%% file: ImplicitCAD_TOG.tex
\documentclass[acmtog,screen,authorversion]{acmart}

\acmSubmissionID{0433}
\AtBeginDocument{%
  \providecommand\BibTeX{{%
    \normalfont B\kern-0.5em{\scshape i\kern-0.25em b}\kern-0.8em\TeX}}}

\setcopyright{acmcopyright}
\acmJournal{TOG}
\acmYear{2022}
\acmVolume{41}
\acmNumber{6}
\acmArticle{276}
\acmMonth{12}
\acmDOI{10.1145/3550454.3555502}

\usepackage{my}

\begin{document}

\title{Implicit Conversion of Manifold B-Rep Solids by Neural Halfspace Representation}

\author{Hao-Xiang Guo}
\email{ghx17@mails.tsinghua.edu.cn}
\affiliation{%
  \institution{Tsinghua University}
  \city{Beijing}
  \country{P~.R~.China}
}
\author{Yang Liu}
\email{yangliu@microsoft.com}
\affiliation{%
  \institution{Microsoft Research Asia}
  \city{Beijing}
  \country{P.~R.~China}
}
\author{Hao Pan}
\email{haopan@microsoft.com}
\affiliation{%
  \institution{Microsoft Research Asia}
  \city{Beijing}
  \country{P.~R.~China}
}

\author{Baining Guo}
\email{bainguo@microsoft.com}
\affiliation{%
  \institution{Microsoft Research Asia}
  \city{Beijing}
 \country{P.~R.~China}
}
\authorsaddresses{Hao-Xiang Guo (work done during internship at Microsoft Research Asia), ghx17@mails.tsinghua.edu.cn;  Yang Liu (corresponding author), Hao Pan, Baining Guo, \{yangliu,haopan,bainguo\}@microsoft.com. }

\renewcommand{\shortauthors}{Guo, et al.}

\begin{abstract}
  \input{src/abstract}
\end{abstract}



\begin{teaserfigure}
  \input{figures/teaser_424}
  \Description{Algorithm illustration of converting a manifold B-Rep solid model to the neural halfspace representation, and sample applications supported by the neural halfspace representation.}
\end{teaserfigure}

\maketitle
\input{src/introduction}

\input{src/relatedwork}

\input{src/overview}
\input{src/NHR}

\input{src/buildtree}

\input{src/method}

\input{src/results}

\input{src/application}

\input{src/conclusion}
\bibliographystyle{ACM-Reference-Format}
\bibliography{src/reference}
\input{src/appendix}

\end{document}

%% file: src/abstract.tex
We present a novel implicit representation --- \emph{neural halfspace representation} (NH-Rep), to convert manifold B-Rep solids to implicit representations. NH-Rep is a Boolean tree built on a set of implicit functions represented by the neural network, and the composite Boolean function is capable of representing solid geometry while preserving sharp features. We propose an efficient algorithm to extract the Boolean tree from a manifold B-Rep solid and devise a neural network-based optimization approach to compute the implicit functions.
We demonstrate the high quality offered by our conversion algorithm on ten thousand manifold B-Rep CAD models that contain various curved patches including NURBS, and the superiority of our learning approach over other representative implicit conversion algorithms in terms of surface reconstruction, sharp feature preservation, signed distance field approximation, and robustness to various surface geometry, as well as a set of applications supported by NH-Rep.

%% file: figures/teaser_424.tex
\begin{overpic}[width=\textwidth]{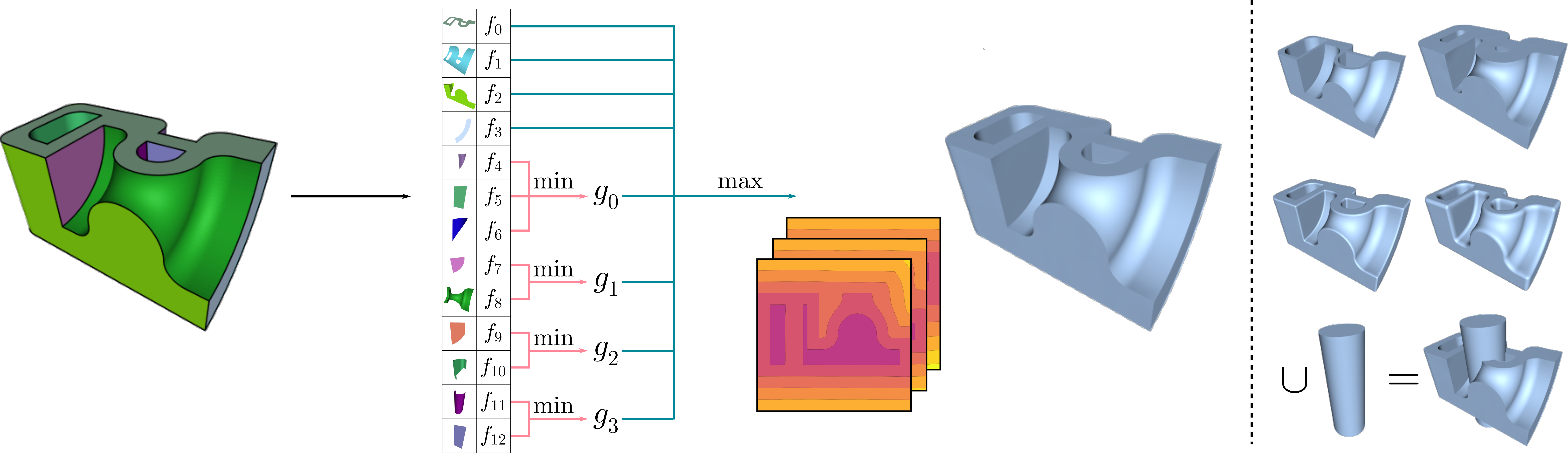}
  \put(7,8){\small $\mS$}
  \put(20,17){\scriptsize NH-Rep}
  \put(19.5,15){\scriptsize conversion}
  \put(52,16){$h(\mx)$}
  \put(66,23){$h(\mx)=0$}
  \put(88,19){\scriptsize sharp offset}
  \put(87,9.3){\scriptsize feature blending}
  \put(89,0){\scriptsize CSG}
\end{overpic}
\caption{\textbf{Left}: Conversion of a manifold B-Rep solid model $\mS$ to the neural halfspace representation. NH-Rep is a function Boolean tree, in which each leaf node corresponds to a B-Rep (sub)patch and associates with a neural implicit function $f_i$.
  The zero isosurface of the Boolean expression $h(\mx)$ at the tree root separates the interior and the exterior space of $\mS$ and represents the boundary surface of $\mS$ and its sharp features faithfully. $h(\mx)$ also approximates the signed distance field of $\mS$, as seen from the sliced images of $h(\mx)$. \textbf{Right}: Sample applications supported by NH-Rep: sharp shape offsetting using different offset distances (top); feature blending using different blending radii (middle); CSG operation on two NH-Reps (bottom).
}
\label{fig:pipeline}

%% file: src/introduction.tex
\section{Introduction} \label{sec:intro}
Implicit solid representations such as constructive solid geometry (CSG) and halfspace representations have gained popularity in computer graphics and CAD/CAM due to many desirable characteristics: it is easy to perform inside/outside queries; it is unbreakable in the sense that challenging operations (\eg Boolean, round, offset) can be applied without numerical failure; they support field-driven design that integrates simulation and manufacturing conditions into shape optimization, thus significantly improving the design to manufacturing cycle consistency \cite{nTopoWhitePaper}; and it is suitable for leveraging fast-growing multicore CPU and GPU architectures for parallelizable computation. Their ability to model signed distance fields is also demanded by many applications~\cite{chen_et_al:DagRep}.  However, existing CAD solid models are often in the form of boundary representations (B-Reps), in which the solid boundary is composed of a set of parameterized surface patches, such as NURBS. To benefit from the aforementioned merits of implicit solid representations, it is necessary to convert B-Reps to usable implicit representations with high fidelity.

An early attempt to convert B-Rep solids to halfspace CSG models is limited to convex solids and B-Reps composed of linear and quadratic patches with linear boundary edges~\cite{Shapiro1993}. High-order solid patches and nonlinear boundary edges are not supported due to their complexity. Inverse CSG approaches, such as Du \etal's work~\shortcite{Du2018}, are restricted by the limited types of solid primitives; thus, it is not easy to represent free-form geometry and sharp features.  Classic implicit reconstruction methods, such as Poisson reconstruction~\cite{Kazhdan2006} and recent neural implicit methods, such as DeepSDF~\cite{Park2019}, lack the ability to faithfully represent sharp features.
The requirements of supporting freeform geometry, preserving sharp features, and modeling SDFs, are still challenges for the implicit conversion of B-Reps.

To address the above challenges, we propose a novel halfspace representation ---
\emph{Neural Halfspace Representation}, abbreviated to \emph{NH-Rep}, is defined by a Boolean tree-based hierarchy of implicit functions, where the implicit function at a leaf node corresponds to a B-rep (sub)patch, and all implicit functions are modeled via a multilayer perceptron (MLP) network.
The composite function $h(\mx): \mx \in \mathbb{R}^3 \longmapsto \mathbb{R}$ by NH-Rep is an implicit representation of a manifold B-Rep solid. Here, the specific construction of the Boolean tree offers the ability to model sharp features in an implicit form, and the use of MLP to represent implicit functions provides a great fitting capability to various surface geometry. NH-Rep has nice properties: (1) the zero isosurface $h(\mx)=0$ separates the interior and exterior of the input B-Rep solid; (2) $h(\mx)=0$ fits the freeform geometry of the input B-Rep solid tightly with normal agreement and sharp feature preservation; (3) the SDF of the B-Rep solid can be approximated by $h(\mx)$ with good quality, by imposing the eikonal equation on $h(\mx)$. An example is illustrated in \cref{fig:pipeline}-left.

Based on the theoretical proof of the existence of NH-Rep for a given manifold B-Rep solid, we develop a method for computing NH-Reps, including an efficient Boolean tree construction algorithm and a learning approach to compute implicit functions. We performed a large-scale benchmark on \myint{10000} manifold B-Rep CAD models to validate the efficacy and robustness of our method. We verified the superiority of NH-Rep over other representative implicit conversion and reconstruction methods in terms of solid approximation fidelity, sharp feature preservation, and robustness to various solid geometries.  Our code is available at \url{https://github.com/guohaoxiang/NH-Rep} for facilitating future research and applications.

%% file: src/relatedwork.tex
\section{Related Work} \label{sec:related}

\paragraph{Solid representation} Boundary representation (B-Rep) and implicit representation (Imp-Rep) are two common solid representations~\cite{shapiro2002solid}. B-Rep represents a solid as a collection of surface, curve, and point elements. The surface and curve elements can be in parametric forms like NURBS, or polygonal forms like triangle facets and polylines.
It is easy to model and edit B-Rep models due to their explicit formulation. Imp-Reps such as Constructive Solid Geometry (CSG)~\cite{ricci1973constructive} and Function Representation (F-Rep) \cite{pasko1995function} provide a constructive modeling scheme and possess the advantages mentioned in \cref{sec:intro}. The former applies Boolean operations to primitive solids; the latter can employ more general R-functions~\cite{shapiro2007semi} based on halfspace functions for modeling and is also suitable for modeling complex shapes in a procedural manner. Our NH-Rep belongs to the category of F-Rep.

\paragraph{Representation conversion} To take advantage of both B-Rep and implicit representations, various applications require conversion between these two representations.
The study of converting implicit representations to B-Reps is mature and many techniques are available, such as polygonalization~\cite{marchingcube} and parameterization of algebraic surfaces. However, the opposite conversion is challenging. The difficulties of converting a B-Rep to a halfspace representation were explored by \cite{Shapiro1993,shapiro1991efficient,SHAPIRO19914,buchele2003three}: additional \emph{separating halfspaces} are required in many circumstances. Shapiro \etal convert B-Reps bounded with linear and quadratic patches by using \emph{linear separators}, but cannot handle high-order patches and nonlinear patch edges. For B-Rep models consisting of simple primitives such as cubes, spheres, and cylinders, a series of CSG inversion works are proposed: binary optimization~\cite{Wu2018}, program synthesizer~\cite{Du2018,xu2021inferring}, evolutionary algorithm~\cite{Friedrich2019}, and learning from data~\cite{Sharma2018,UCSG-Net,ren2021csg,yu2021capri}; but they cannot accurately represent freeform geometry due to their restricted primitive types.
Boundary-sampled halfspace (BSH)~\cite{Du2021} defines the shape as sparsely placed samples on the halfspace boundaries and provides greater agility and expressiveness than CSG for shape modeling, and it also simplifies the conversion process but has challenges in handling tangentially contacted surface patches, which are common in CAD models. A simple failure case is provided in \cref{appendix:bsh}.

\paragraph{Reconstruction-based implicit conversion}
By treating a B-Rep model as a collection of densely sampled points, many implicit-based reconstruction techniques~\cite{Berger2017,Carr2001,Ohtake2003} can be used to convert point clouds into implicit representations.  For recovering shape features, Kazhdan \etal~\shortcite{Kazhdan2013} use an indicator field to represent a 3D shape and approximate features by penalizing the difference between the surface gradients and the oriented point normals. Oztireli \etal~\shortcite{Oztireli2009} uses implicit moving least squares (MLS) to represent 3D shapes while estimating adaptive functional weights to handle features and outliers. However, these methods cannot model $C^1$ discontinuous features accurately, as the underlying surface representation is globally smooth, not piecewise smooth.   Recent neural implicit approaches~\cite{Chen2019,Park2019,Peng2020,Chabra2020,Chibane2020,Jiang2020,liu2021deep,yang2021geometry} represent the shape geometry as a signed distance field or an occupancy field, and some works~\cite{Gropp2020,sitzmann2019siren,williams2021neural} overfit the input point cloud to achieve high accuracy, but the learned function is either piecewise linear when using MLPs with ReLU~\cite{lei2020analytic} or globally smooth; therefore, it cannot model (curved) sharp features faithfully.  The works of BSP-Net~\cite{bspnet} and CVXNet~\cite{Deng2020} infer a Boolean CSG tree of a set of planes to represent a 3D shape in a simple and compact form. However, their approximation quality to the input is restricted by the plane geometry and the number of planes. By utilizing the B-Rep patch information explicitly and composing the learned neural implicit functions with a Boolean tree, we obtain faithful implicit conversion with sharp feature preservation.

\paragraph{Neural B-Rep representation} The complex data structure of B-Rep representations poses a challenge in integrating B-Rep with modern machine learning techniques. Several approaches have been developed to address this challenge. UV-Net~\cite{uvnet} converts B-Rep to a face-adjacency graph in which the face and edge correspond to a parametric surface and curve, respectively. 2D CNN and 1D CNN are performed in the UV domain of each face and edge, and the learned features are aggregated by graph convolution. BRepNet~\cite{lambourne2021brepnet} learns the features on B-Rep faces, edges, loops, coedges, and vertices by graph convolution. These works are mainly designed for shape analysis tasks, such as classification and segmentation. Recently Willis \etal~\shortcite{willis2021joinable} associate the one-hot feature (element type) for B-Rep faces and edges, and pass messages through the B-Rep graph to predict the joint information for part assembly. Wu \etal~\shortcite{deepcad} proposes a generative network to infer a series of CAD commands to create or reconstruct CAD models. However, its CAD reconstruction capability from point clouds is limited by the model complexity and the predefined commands. Our approach utilizes the patch adjacency of B-Rep to deduce the Boolean tree for converting B-Rep to implicit representation and has good scalability to deal with complex models.

\paragraph{Feature-sensitive distance fields} Storing distance field values in discrete grids~\cite{jones20063d} is a popular way to convert explicit models to implicit representation. Adaptive octree~\cite{Frisken2000} and hp-refinement~\cite{koschier2016hierarchical} are common techniques to make storage compact while maintaining sufficient accuracy. To preserve sharp corners and sharp features, additional information and operations are required to store, such as \emph{directed distances} in the $x$-, $y$- and $z$-direction~\cite{Kobbelt2001}, exact intersection points and normals~\cite{Ju2002}, the nearest triangle information of the grid points to the input mesh~\cite{huang2001complete}, offset distance fields~\cite{qu2004feature}, and density gradient~\cite{10.2312:VG:VG05:109-116}. To represent high-quality surface details and curved sharp features, high-resolution (adaptive) grids are needed in all these approaches and take up large storage spaces. In contrast, the neural implicit representation can approximate the signed distance field and achieves a good balance between storage efficiency and approximation accuracy; furthermore, the Boolean operations of NH-Rep ensure faithful feature preservation.

%% file: src/overview.tex
\section{Overview} \label{sec:overview}
Our paper is organized as follows. In \cref{subsec:NHR}, we first introduce the terminologies of boundary representation and halfspace representation, then present our solution for converting manifold B-Rep solids to halfspace representation --- Neural Halfspace Representation (NH-Rep), which is built on a special Boolean tree of a set of implicit functions. We develop a Boolean tree construction algorithm (\cref{subsec:tree}) and a neural network-based optimization algorithm to determine the implicit functions associated with the Boolean tree (\cref{subsec:learning}). In \cref{sec:result}, we provide extensive experimental analysis and ablation studies to verify the efficiency and superiority of our approach. In \cref{sec:app}, we demonstrate a series of applications supported by NH-Rep.

%% file: src/NHR.tex
\section{Neural Halfspace Representation} \label{sec:NHR}
\subsection{Boundary representation and halfspace representation} \label{subsec:background}

\paragraph{Boundary representation}
A B-Rep solid is a 3D volume surrounded by a collection of non-overlapped manifold surface patches. For a solid $\mS$ with $L$ patches, we denote these patches as $\mP_1, \ldots, \mP_L$, and the boundary of $\mS$ as $\partial \mS$ which satisfies $\partial \mS = \bigcup_i \mP_i$.  Any surface patch of a B-Rep solid can be in the form of a parametric surface or a set of polygonal facets. Without loss of generality, we assume that all surface patches have a consistent normal orientation, pointing outside of the volume. For a B-Rep solid $\mS$,  $\partial \mS, \mathcal{I}(\mS)$ and $\mathcal{E}(\mS)$ denote the boundary surface, the interior space and the exterior space of $\mS$, respectively. For convenience, we define $\overline{\mathcal{I}}(\mS) =  \mathcal{I}(\mS) \bigcup \partial \mS$. In our paper, we assume that $\partial \mS$ is manifold and the solid volume is not empty.

\input{figures/solid}
\paragraph{Feature curves of B-Rep} The boundary of a B-Rep patch is formed by a set of 3D curves, each of which is shared by two adjacent B-Rep patches. We call these curves \emph{feature curves}. The endpoints of a feature curve are called \emph{feature corners} and the other points on the feature curve are called \emph{feature points}. We call a feature point $\mp$ \emph{convex}, if the interior dihedral angle at $\mp$, denoted by $\gamma_\mp$, is smaller than \ang{180}; \emph{concave} if $\gamma_\mp > \ang{180}$; and \emph{smooth} if $\gamma_\mp = \ang{180}$. ``Convex'' or ``concave'' means that $\partial \mS$ is only $C^0$ smooth at $\mp$. Here $\gamma_\mp$ is called \emph{feature angle} at $\mp$. We call a feature curve $\mC$ \emph{convex} if all feature points on $\mC$ are not concave;  \emph{concave} if all feature points on $\mC$ are not convex;  \emph{smooth} if all feature points on $\mC$ are smooth; \emph{hybrid} if $\mC$ contains both convex and concave feature points. We say a feature curve $\mC$ \emph{sharp} if $\max_{\mp \in \mC} |\ang{180}-\gamma_\mp| \geq \delta_s$, $\delta_s > 0$ is the user-defined sharp angle threshold.  The right inset illustrates a B-Rep solid with different types of feature curves: convex in black, concave in red, and smooth in blue.

\input{figures/featurefig}

\paragraph{Halfspace representation}
An implicit function $f: \mx \in \mathbb{R}^3 \mapsto \mathbb{R}$ defines a halfspace: $\{\mx \in \mathbb{R}^3: f(\mx) \leq 0\}$. For simplicity, we abbreviate the notation as $f \leq 0$. Boolean operations (union, intersection, subtraction) can be defined over halfspaces using  $\bm{\max}$ and $\bm{\min}$ functions as follows. The union of a set of halfspaces $\{f_i \leq 0\}_{i=1}^m$ is $\{\mx\in \mathbb{R}^3: \bm{\min}\bigl(f_1(\mx), \cdots, f_m(\mx)\bigr) \leq 0 \}$; their intersection is $\{\mx\in \mathbb{R}^3: \bm{\max}\bigl(f_1(\mx), \cdots, f_m(\mx)\bigr) \leq 0 \}$; the difference of two halfspaces $f \leq 0$ and $g \leq 0$ is $\{\mx\in \mathbb{R}^3: \bm{\max}\bigl( f(\mx), -g(\mx) \bigr) \leq 0 \}$.  By compositing some Boolean operations on halfspaces, a \emph{Boolean expression}, or called \emph{CSG expression}, is obtained.  A Boolean expression can be rewritten in \emph{Boolean tree} format: (1) each leaf node stores an implicit function; (2) each non-leaf node stores a Boolean operation: $\bm{\max}$ or $\bm{\min}$, which is applied to its child nodes; (3) the composite function at the root node, denoted by $h(\mx)$, is the Boolean expression. For convenience, we denote the Boolean operation on the non-leaf tree node $\mr$ as $\bm \op(r)$, and use $\bm\op^+(\mr)$ to denote the opposite operation of $\bm\op(\mr)$, \ie  $\bm\op^+(\mr) := \bm{\max}$, if $\bm \op(\mr) = \bm\min$;  $\bm\op^+(\mr) := \bm{\min}$ if $\bm \op(\mr) = \bm\max$.

A Boolean tree is a halfspace representation of a B-Rep solid if its composite function at the tree root can classify the point membership of $\mS$ correctly,
\ie holding the following properties:
\begin{equation} \label{eq:classification}
    \begin{cases}
        h(\mx) < 0, & \forall \mx \in \mathcal{I}(\mS); \\
        h(\mx) > 0, & \forall \mx \in \mathcal{E}(\mS); \\
        h(\mx) = 0, & \forall \mx \in \partial\mS.
    \end{cases}
\end{equation}
\cref{fig:convexconcave}-(b) provides a Boolean tree example that can represent a 2D B-Rep solid shown in \cref{fig:convexconcave}-(a).
In \cref{fig:convexconcave}-(c,d,e), the zero isocurves of the Boolean expressions on the tree nodes are also illustrated.

\paragraph{B-Rep to halfspace representation}
Any surface patch of a B-Rep solid can be represented as a subset of the zero isosurface of an implicit function. All implicit functions associated with B-Rep patches define a set of halfspaces. For a solid $\mS$ with $L$ patches, all halfspaces partition $\mathbb{R}^3$ into $2^L$ cells (some are possibly empty).
The describability theorem~\cite{Shapiro1993} states that there exists a Boolean expression on these halfspaces that can represent $\mS$ if and only if for each nonempty cell, all points in the cell have the same point membership classification with respect to the solid volume.
This condition implies \cref{eq:classification}, but is not easy to satisfy in general, except for convex solids. Shapiro and Vossler~\shortcite{Shapiro1993} proposed adding \emph{separators}, \ie additional halfspaces, to partition those cells violating the condition until the mixed regions are separated. Linear separators for B-Rep solids composed of linear and quadratic patches were studied in their work. Nonlinear separators are essential for complicated B-Rep solids but their constructions are unknown. Our work provides an explicit way to compute the Boolean expression and the halfspace functions for general B-Rep solids and introduces nonlinear separators via patch decomposition (\cref{subsec:treeconstruction}) when necessary.

\subsection{Neural halfspace representation}\label{subsec:NHR}
We present a special halfspace representation --- \emph{neural halfspace representation} to convert an arbitrary B-Rep solid to an implicit presentation. To facilitate the definition of \emph{neural halfspace representation}, we first introduce the following useful terminologies.

\begin{definition}
    For a B-Rep solid $\mS$ with $L$ patches,  a set of subpatches $\{\ms_1, \ldots, \ms_n\}$ is called a \emph{decomposed patch set} of $\mS$ if (1) $\bigcup_{i=1}^n \ms_i = \partial \mS$; (2) $\forall i, \emptyset \neq \ms_i \subseteq \mP_k, \exists k \in \{1, \cdots, L\} $; (3) $\area(\ms_i \cap \ms_j) = 0, \forall i \neq j$.
\end{definition}

\begin{definition}
    A Boolean tree $\mathcal{T}$ built on a set of implicit functions $\{f_1, \cdots, f_n\}$ is called a \emph{patch-based Boolean tree}, if there exists a decomposed patch set $\{\ms_1, \ldots, \ms_n\}$ of $\mS$ such that $\ms_i$ lies on the zero surface of $f_i$, \ie $\ms_i \subseteq \{\mx \in \mathbb{R}^3: f_i(\mx) = 0\}, \forall i$.
\end{definition}

The following theorem guarantees that any manifold B-Rep solid can be converted to a halfspace representation.
\begin{theorem}
    For a B-Rep solid $\mS$, there exists a patch-based Boolean tree $\mathcal{T}$ to represent $\mS$ in half-space representation, \ie the Boolean expression of $\mathcal{T}$ satisfies \cref{eq:classification}.
\end{theorem}
The proof of the above theorem includes two parts, sketched as follows. We first provide an explicit way to build the tree structure of $\mathcal{T}$ and determine the decomposed patch set explicitly, then prove that there exists a set of implicit functions of $\mathcal{T}$ whose composition function at the root node fulfills \cref{eq:classification}. We detail the first part in \cref{subsec:tree} and leave the proof to \cref{appendix:proof}.

With the guarantee provided by the above theorem and the built patch-based Boolean tree structure, the computation of implicit functions of $\mathcal{T}$ is formulated as an optimization problem. To maximize the fitting ability of implicit functions to various and complex B-Rep patch geometry, we represent all implicit functions by a neural network, which maps a 3D point to an $n$-dimensional vector in which each entry is the value of an implicit function evaluated at the given point. The optimization details are presented in \cref{subsec:learning}.

As we use neural networks to represent implicit functions of the patch-based Boolean tree, we name the halfspace representation under this setup by \emph{neural halfspace representation}.

%% file: figures/solid.tex
\begin{wrapfigure}[8]{r}{0.26\columnwidth}
    \includegraphics[width=0.25\columnwidth]{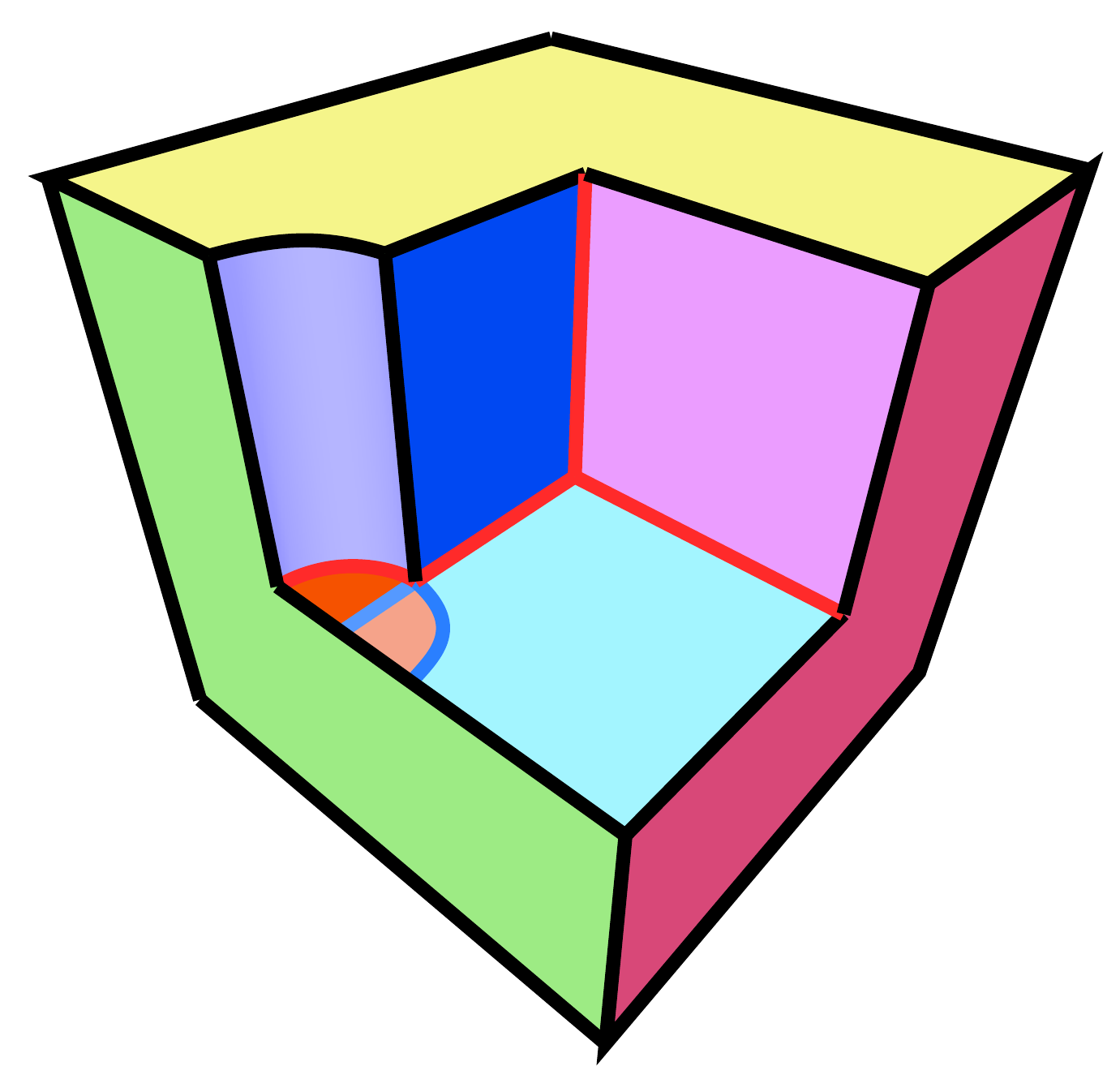}
    \Description{Illustration of a B-Rep solid model. B-Rep patches are rendered in different colors, and feature curves are highlighted: convex in black, concave in red, smooth in blue.}
\end{wrapfigure}

%% file: figures/featurefig.tex
\begin{figure*}[t]
    \centering
    \begin{overpic}[width=\linewidth]{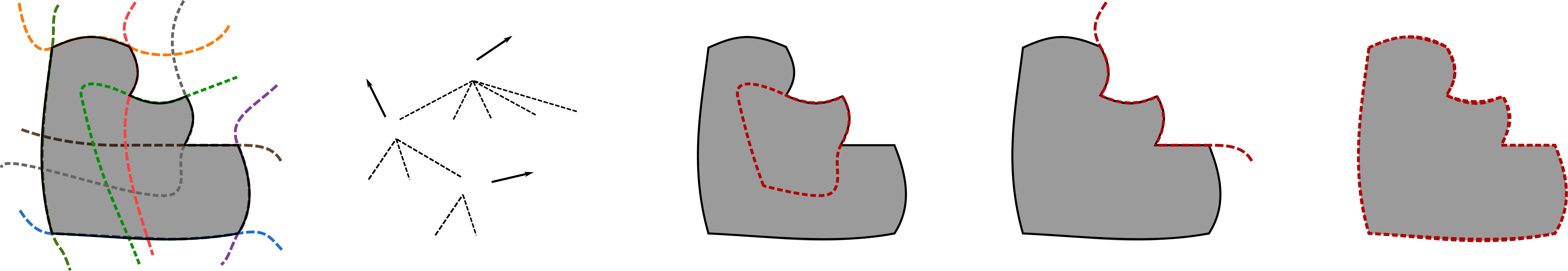}
        \put(9,-1.5){\small \textbf{(a)}}
        \put(30,-1.5){\small \textbf{(b)}}
        \put(50,-1.5){\small \textbf{(c)}}
        \put(70,-1.5){\small \textbf{(d)}}
        \put(91,-1.5){\small \textbf{(e)}}
        \put(52,12.5){\scriptsize $g_1(\mx)=0$}
        \put(72,12.5){\scriptsize $g_2(\mx)=0$}
        \put(94,12.5){\scriptsize $h(\mx)=0$}
        \put(0.5,14.5){\scriptsize $c_1$}
        \put(2.5,17.3){\scriptsize $c_2$}
        \put(0.5,2.8){\scriptsize $c_3$}
        \put(12.8,0){\scriptsize $c_4$}
        \put(17.9,7.5){\scriptsize $c_5$}
        \put(10.4,17.3){\scriptsize $c_6$}
        \put(15.3,12){\scriptsize $c_7$}
        \put(7.4,17.3){\scriptsize $c_8$}

        \put(4.5,13.5){\scriptsize $\mP_1$}
        \put(3,10){\scriptsize $\mP_2$}
        \put(5,2.9){\scriptsize $\mP_3$}
        \put(14,4.3){\scriptsize $\mP_4$}
        \put(12.8,6.8){\scriptsize $\mP_5$}
        \put(10.3,8.9){\scriptsize $\mP_6$}
        \put(8.7,9.6){\scriptsize $\mP_7$}
        \put(7,12.5){\scriptsize $\mP_8$}
        \put(33,15){\scriptsize $h$}
        \put(22.5,13){\scriptsize $g_2$}
        \put(34.5,6){\scriptsize $g_1$}

        \put(28.9,12.3){\scriptsize$\bm{\max}$}
        \put(28.2,5){\scriptsize$\bm{\max}$}

        \put(24.5,8.6){\scriptsize$\bm{\min}$}
        \put(28.3,8.6){\scriptsize $f_1$}
        \put(30.7,8.6){\scriptsize $f_2$}
        \put(33.5,8.6){\scriptsize $f_3$}
        \put(36.3,8.6){\scriptsize $f_4$}

        \put(23,4.6){\scriptsize $f_5$}
        \put(25.5,4.6){\scriptsize $f_8$}
        \put(27.5,1){\scriptsize $f_6$}
        \put(29.8,1){\scriptsize $f_7$}

    \end{overpic}
    \caption{\textbf{(a)}: A 2D B-Rep solid (shaded region) surrounded by eight curve segments $\{\mP_i\}_{i=1}^8$. $\mP_i$ corresponds to an implicit function $f_i$ whose zero isocurve $c_i$ is depicted as a dash line. \textbf{(b)}: The Boolean tree of \textbf{(a)}. $h=\bm{\max}\bigl(f_1, f_2, f_3, f_4, \bm\min\bigl(f_5, f_8, \bm\max(f_6, f_7)\bigr)\bigr)$. The red curves in \textbf{(c)}, \textbf{(d)}, and \textbf{(e)} are the isocurves of the Boolean expression $g_1$, $g_2$ and $h$, respectively. $h(\mx)=0$ recovers the boundary of the B-Rep solid exactly. }
    \Description{Illustration of a Boolean tree of a 2D B-Rep solid.}
    \label{fig:convexconcave} \vspace{-4mm}
\end{figure*}

%% file: src/buildtree.tex
\section{Boolean tree structure construction} \label{subsec:tree}

\subsection{Algorithm overview}
The design of Boolean tree structures is based on the following intuitive idea. Notice that the Boolean operations on halfspaces have clear geometry meaning: union, intersection, and subtraction; we propose to partition the space via these operations defined over halfspaces progressively, until one of the partition cells matches the volume of the input B-Rep, \ie satisfying the describability theorem~\cite{Shapiro1993}.

We sketch the above idea as follows, using a 2D example shown in \cref{fig:creation}. The initial partition space, denoted by $S_p$, is the full space. We first select a maximal group of patches, each of which shares convex feature curves with its neighbor patches, \ie $c_1, c_2, c_3, c_4$.
We assume that by defining the appropriate implicit functions for these selected patches, the intersection of their halfspaces can contain $\overline{\mathcal{I}}(\mS)$. We update $S_p$ with this intersected region. We then select a maximal group of patches from the rest of the unprocessed patches,
each of which shares concave feature curves with its unprocessed neighbor patches, \ie $c_5, c_8$.
We also assume that, by defining the appropriate implicit functions on $c_5, c_8$, the union of their corresponding halfspaces can be intersected with $S_p$, and these patches become part of the boundary of the intersected region. $S_p$ is then updated by this new intersected region. Union and intersection operations can be executed alternatively until all surface patches are processed. Finally, $S_p$ matches the volume of the input B-Rep solid, as $S_p$ is tightly surrounded by all B-Rep patches. \cref{fig:creation}-(b-d) illustrates each step. In the next subsection, we turn the above idea into a rigorous and executable algorithm.

\input{figures/creation}

\input{figures/alg}

\subsection{Tree structure construction} \label{subsec:treeconstruction}

For a B-Rep solid $\mS$, we provide a recursive approach to construct a patch-based Boolean tree. It includes two steps: \emph{tree initialization} and
\emph{tree node creation}, and it relies on the following graph structure.

\paragraph{Patch graph} We define an undirected multigraph according to the adjacency of all surface patches of $\mS$. Each \emph{graph vertex} corresponds to a surface patch;  each \emph{graph edge} corresponds to a feature curve shared by two adjacent patches, and each \emph{graph edge} $\me$ is labeled with its corresponding feature curve type, denoted by $\texttt{type}(\me)$.  This multigraph is called \emph{patch graph}, denoted by $\mG$. \cref{fig:algorithm}-(b) depicts the patch graph of the B-Rep model shown in \cref{fig:algorithm}-(a).

\paragraph{Tree initialization}
If $\mG$ contains multiple \emph{maximal connected subgraphs} (in short, MC subgraph), \ie the volume of $\mS$ contains multiple disconnected components, a tree root node $\mr$ is created, and each MC subgraph will be used to create child nodes under the root.  The Boolean operation on $\mr$ is set to $\bm\min$, to combine all components.
If $\mG$ contains only one MC subgraph, we set $\mr$ as a virtual node: $\mr =\emptyset$ and $\op(\mr) =\bm\min$. The virtual node does not appear in the final Boolean tree structure.

We take each MC-subgraph and the tree root as input to the following recursive tree node creation algorithm.

\paragraph{Tree node creation}
This step takes an MC-subgraph $\mG'$ and a parent tree node $\mr$ as input. From $\mG'$, we select graph vertices that have no connected graph edges with label $\texttt{edgetype}(\bm\op^+(\mr))$, to form a vertex set $\mQ$. Here, we define $\texttt{edgetype}(\bm{\max}) := \texttt{convex}$, $\texttt{edgetype}(\bm{\min}) := \texttt{concave}$.
As the graph vertex is dual to a surface patch, this selection does the following job: (1) when $\op(\mr) = \bm\min$, $\mQ$ collects those surface patches in $\mG'$ that have no adjacent patches in $\mG'$ or only share convex or smooth feature curves with other adjacent patches in $\mG'$; (2) when $\op(\mr) = \bm\max$, $\mQ$ collects those surface patches in $\mG'$ that have no adjacent patches in $\mG'$ or only share concave or smooth feature curves with other adjacent patches in $\mG'$.
The formed $\mQ$ is processed as follows.

\vspace{2pt}
\textbf{Case 1: $\mQ = \emptyset$}. There are two subcases that $\mQ$ can be empty.  The first subcase occurs when $\mr$ is the root node associated with $\op(\mr) =\bm\max$ operation, but $\mS$ does not contain any convex feature edges. For this special subcase (depicted in \cref{fig:allconcave}), we only need to set $\op(\mr) =\bm\min$ and construct the Boolean tree directly.  The second subcase occurs when any patch in $\mG'$ has both convex and concave features shared with adjacent patches. \cref{fig:repair}-(b) depicts such an MC-subgraph in this situation.  For this subcase, we randomly select one patch from $\mG'$, denoted by $\mP$, and decompose it into a set of subpatches, such that the boundary curves of any subpatch do not contain convex and concave feature curve segments of $\mP$ simultaneously. We call this step \emph{patch decomposition}. \cref{fig:repair}-(c) illustrates the patch decomposition step. We replace $\mP$ with these subpatches and update $\mG'$ accordingly. Here, any graph edge connecting with two adjacent subpatches is labeled \emph{smooth}. We then recompute $\mathcal{Q}$ from the updated $\mG'$ and execute the tree node creation algorithm.

\input{figures/allconcave}
\input{figures/repair}

\vspace{2pt}\textbf{Case 2: $\mathcal{Q} \neq \emptyset$}.
A child node $\mr'$ is created under $\mr$, and the Boolean operator at $\mr'$ is set to the opposite operation of $\mr$, \ie $ \bm \op(\mr') :=\, \bm\op^+(\mr)$. For each graph vertex $v \in \mathcal{Q}$, a child node is created under $\mr'$. These added child nodes are tree leaf nodes because each of them corresponds to a surface (sub)patch. $\mG'$ is updated by removing the vertices of $\mQ$ and their connected edges from $\mG'$. Then we process every MC-subgraph in the updated $\mG'$ by feeding it and $\mr$ to our tree node creation algorithm.

The pseudocode of the above recursive node creation algorithm is provided in \cref{alg:csg}.
\cref{fig:algorithm} illustrates how the Boolean tree is created for a 2D B-Rep solid, step by step.

\begin{algorithm}
    \SetKwInOut{Input}{Input}\SetKwInOut{Output}{output}
    \SetAlgoLined
    \Input{patch graph $\mG'$, parent tree node $\mr$}
    \KwResult{updated Boolean tree structure}
    Create $\mathcal{Q}$ \;
    \If{$\mathcal{Q} \neq \emptyset$}
    {
        create child node $\mr'$ under $\mr$ and set $\bm\op(\mr') := \, \bm \op^+(\mr)$\;
        $\forall \mv \in \mathcal{Q}$, create a leaf node under $\mr'$ \;
        graph update: $G' := G' \backslash \mathcal{Q} $\;
        \For{each MC-subgraph $\mG'' \subseteq \mG'$}
        {
            ConstructTreeNode($\mG''$, $\mr'$)\;
        }
    }
    \Else{
        \If{No convex edges in $\mG'$}
        {
            set $\bm\op(\mr') = \min $\;
            $\forall \mv \in \mathcal{Q}$, create a leaf node under $\mr'$\;
        }
        \Else{
            Pick $\mv \in \mG'$, decompose its corresponding patch $\mP$\;
            Update $\mG'$ based on the updated patch layout\;
            ConstructTreeNode($\mG'$, $\mr$)\;
        }
    }
    \caption{ConstructTreeNode} \label{alg:csg}
\end{algorithm}

\paragraph{Termination of tree construction} Since $\mG$ has finite graph vertices and patch decomposition always reduces the number of graph nodes that connect with convex and concave edges, the recursive algorithm ends in finite steps. The subpatches from patch decomposition and the original non-decomposed patches form the final decomposed patch set, and each leaf node corresponds to one of the (sub)patches. The tree node creation step ensures that $\bm\max$ and $\bm\min$ appear alternately in different layers of tree nodes. Thus, the Boolean tree created is a patch-based Boolean tree.

\paragraph{Implementation of patch decomposition} We implement patch decomposition as follows.
For a surface patch in polygonal mesh format, we first employ face-splitting to ensure that no triangle contains feature edge segments with different feature types. Then, on the dual graph of the mesh, we formulate a facet labeling problem such that the facets with the same label do not contain convex and concave feature edges simultaneously. The problem is solved using the Min-Cut algorithm. The labeling result induces different subpatches. For a patch in parametric surface format such as NURBS, we need to discrete it first as a triangular mesh and then compute the subpatches. Here, we record the parametric coordinates of each mesh vertex, so that we can access surface points on the parametric surfaces exactly in the later NH-Rep computation stage, without suffering discretization artifacts. \cref{fig:repair} illustrates patch decomposition on a simple example.

%% file: figures/creation.tex
\begin{figure}[t]
    \centering
    \begin{overpic}[width=\columnwidth]{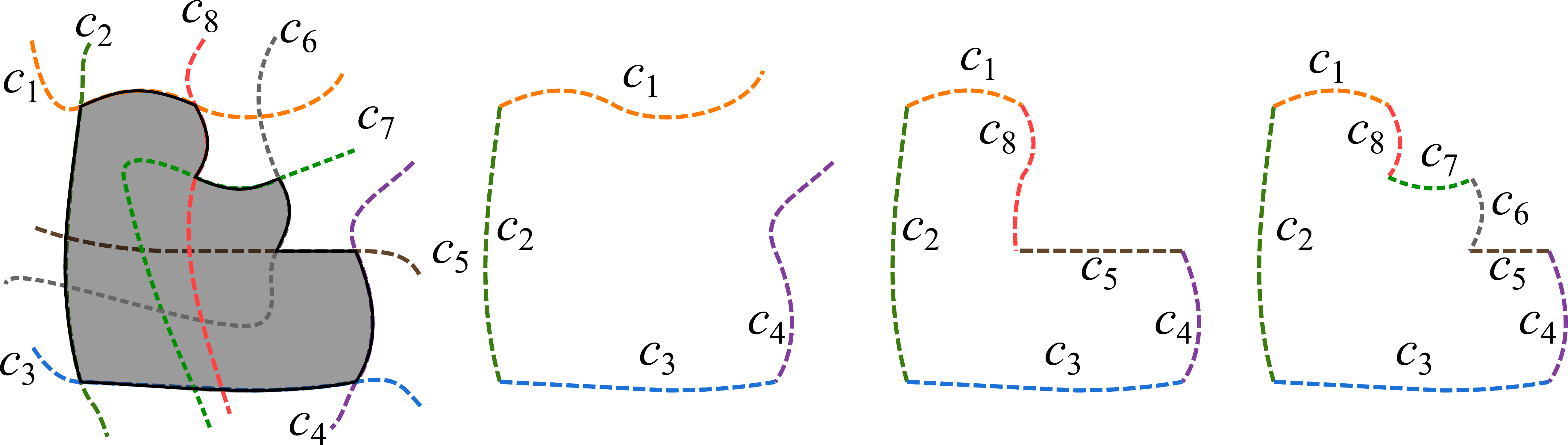}
        \put(12,-1.5){\small \textbf{(a)}}
        \put(37,-1.5){\small \textbf{(b)}}
        \put(63,-1.5){\small \textbf{(c)}}
        \put(86,-1.5){\small \textbf{(d)}}
    \end{overpic}
    \caption{2D illustration of Boolean tree construction. \textbf{(a)}: An input B-Rep solid. The zero isocurves of the implicit functions are illustrated, same as \cref{fig:convexconcave}-(a). \textbf{(b)}-\textbf{(d)} are the intermediate partition results. \textbf{(b)}: The intersection of  $\{f_1\leq 0, f_2\leq 0 ,f_3\leq 0,f_4 \leq 0\}$. The corresponding Boolean expression is $\bm\max(f_1,f_2,f_3,f_4)$. \textbf{(c)}: The union of $f_5 \geq 0$ and $f_8 \leq 0$ is intersected with the space defined in \textbf{(b)}. The corresponding Boolean expression is $\bm\max\bigl(f_1,f_2,f_3,f_4, \bm\min(f_5,f_8)\bigr)$. \textbf{(d)}: The region surrounded by $c_6,c_7$ is added to the partition.  The corresponding Boolean expression is $\bm\max\bigl(f_1,f_2,f_3,f_4, \bm\min\bigl(f_5,f_8, \bm\max(f_6,f_7)\bigr)\bigr)$, yielding the Boolean tree shown in \cref{fig:convexconcave}-(b).}
    \Description{Illustration of Boolean tree construction}
    \label{fig:creation} \vspace{-2mm}
\end{figure}

%% file: figures/alg.tex
\begin{figure*}[t]
  \centering
  \begin{overpic}[width=0.95\linewidth]{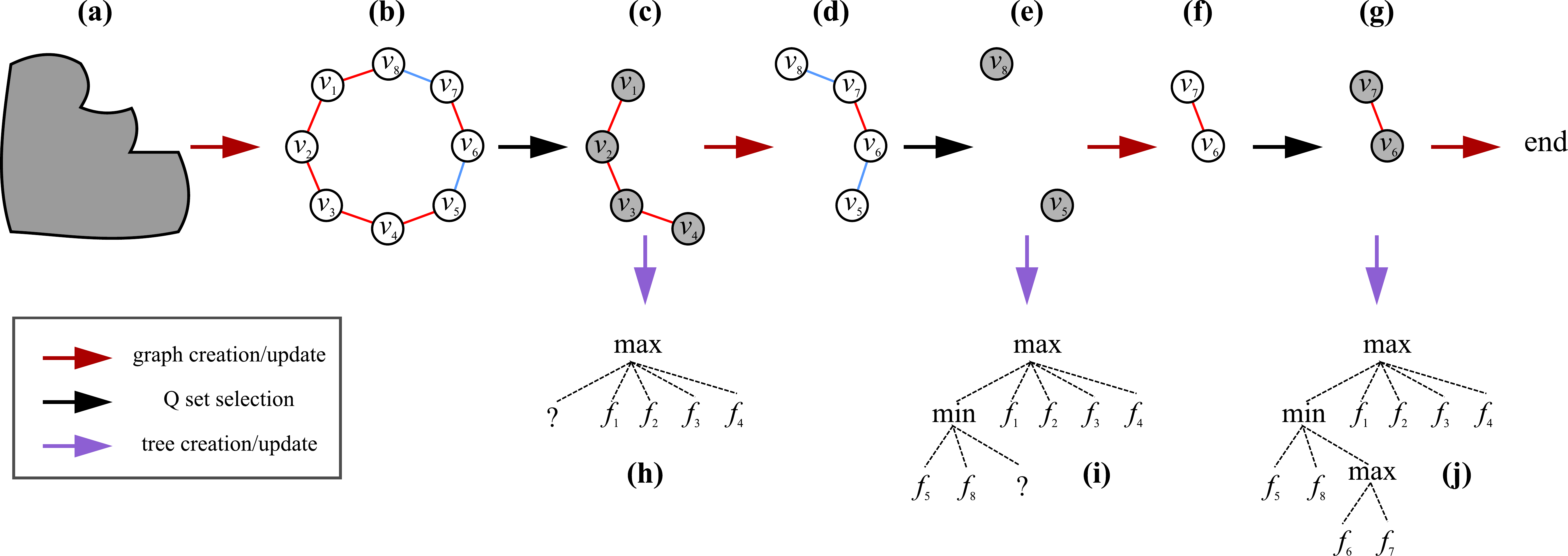}
    \put(2,32.5){\scriptsize $\mP_1$}
    \put(0.2,25){\scriptsize $\mP_2$}
    \put(5,18.5){\scriptsize $\mP_3$}
    \put(12.2,23){\scriptsize $\mP_4$}
    \put(9.5,26.2){\scriptsize $\mP_5$}
    \put(7,27){\scriptsize $\mP_6$}
    \put(6,28.8){\scriptsize $\mP_7$}
    \put(4,30){\scriptsize $\mP_8$}
  \end{overpic}
  \vspace{-5mm}
  \caption{Illustration of Boolean tree construction for a 2D B-Rep solid. \textbf{(a)}: An input 2D B-Rep solid with eight curve patches. A patch graph is created first as shown in \textbf{(b)}. Graph vertex $v_i$ corresponds to patch $\mP_i$, and the color of graph edges encodes the edge types: red (convex), blue (concave). \textbf{(c)}-\textbf{(g)} illustrate the tree node creation step by step. \textbf{(c)}, \textbf{(e)}, \textbf{(g)} are the $Q$ set in each step; \textbf{(d)} and \textbf{(f)} are the updated patch graph by removing vertices of $Q$ created from previous step. \textbf{(h)}, \textbf{(i)} and \textbf{(j)} are the Boolean tree during  construction. $f_i$ corresponds to $v_i$ and $\mP_i$.}
  \Description{Illustration of Boolean tree construction for a 2D B-Rep solid.}
  \label{fig:algorithm} \vspace{-3mm}
\end{figure*}

%% file: figures/allconcave.tex
\begin{figure}[t]
    \centering
    \begin{overpic}[width=0.55\columnwidth]{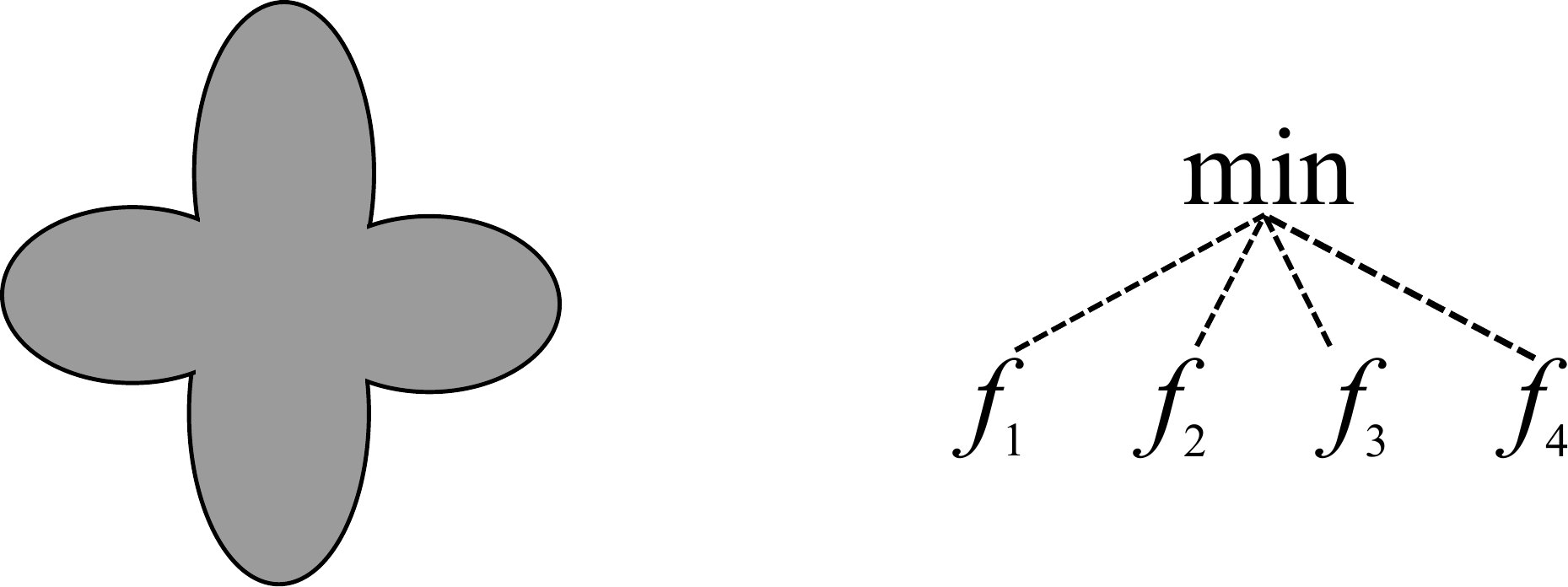}
        \put(16,31){\scriptsize \textbf{$\mP_1$}}
        \put(2,17.5){\scriptsize \textbf{$\mP_2$}}
        \put(16,4){\scriptsize \textbf{$\mP_3$}}
        \put(29,17.5){\scriptsize \textbf{$\mP_4$}}
    \end{overpic} \vspace{-2mm}
    \caption{A 2D B-Rep solid formed by four patches contains concave features only (left). The operation at the root node should be set to $\bm\min$.}
    \label{fig:allconcave} \vspace{-3mm}
\end{figure}

%% file: figures/repair.tex
\begin{figure}
    \centering
    \includegraphics[width=0.85\columnwidth]{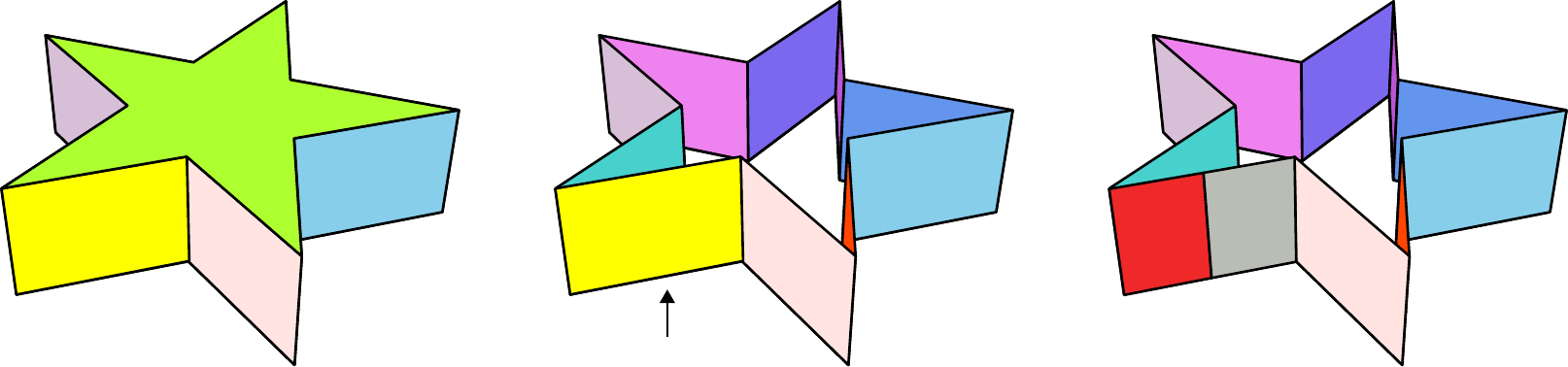}
    \caption{Illustration of patch decomposition.
        \textbf{Left}: A star-shape B-Rep that contains 12 patches (in different colors). \textbf{Middle}: 10 vertical patches form a graph vertex set, triggering the occurrence of the second subcase of $\mathcal{Q} = \emptyset$. The yellow patch is chosen for decomposition. \textbf{Right}: Subpatches obtained by patch decomposition.} \label{fig:repair}
    \vspace{-5mm}
\end{figure}

%% file: src/method.tex
\section{{NH-Rep} computation} \label{subsec:learning}

With a built patch-based Boolean tree structure, our objective is to compute a set of implicit functions $\{f_i\}_{i=1}^n$ such that the conditions of \cref{eq:classification} can be met.
We rewrite \cref{eq:classification} with respect to the decomposition patch set $\{\ms_1, \ldots, \ms_n\}$ as follows.
\begin{align}
  h (\mx)              & = 0,        & \forall \mx \in \ms_i, \forall i; \label{eq:hpos}       \\
  h (\mx)              & > 0,        & \forall \mx \in \mathcal{E}(\mS); \label{eq:hexterior}  \\
  h (\mx)              & < 0,        & \forall \mx \in \mathcal{I}(\mS);  \label{eq:hinterior} \\
  f_i(\mx)             & = 0,        & \forall \mx \in \ms_i, \forall i; \label{eq:fpos}       \\
  \nabla_\mx f_i (\mx) & = \mn(\mx), & \forall \mx \in \ms_i, \forall i. \label{eq:normal}
\end{align}
Here, $\mn(\mx)$ is the oriented surface normal at $\mx$. We impose \cref{eq:fpos,eq:normal} to ensure that the zero surface of $f_i$ contains $\ms_i$ and is the first-order approximation of $\ms_i$.

We design a learning approach to compute the implicit functions that satisfy the above conditions. We use a multilayer perceptron (MLP) with three hidden layers of size $256$
as our network, shown in \cref{fig:network}. It takes a 3D point coordinate $\mx$ as the input and output $n$ values, each of which is the value of $f_i$ at $\mx$.
These $n$ values are passed to the Boolean tree to obtain the composite function value $h(\mx;\theta)$ ($\theta$ denotes the network parameters).
Here, we use \texttt{SoftPlus} as the activation function to ensure that MLPs have sufficient smoothness to represent smooth patches.
The weight of the network is initialized with geometric
initialization~\cite{Atzmon2020}.

\input{figures/network}

Our loss function consists of the following terms.
\paragraph{Position loss} The position loss $E_p$ is derived from \cref{eq:hpos,eq:fpos}:
\begin{equation}
  E_p := \frac{1}{N} \sum_{i=1}^n \sum_{\mx \in \ms_i} |f_i(\mx;\theta)| + |h(\mx;\theta)|.
  \label{eq:pos_loss}
\end{equation}
Here, $N$ is the total number of sample points.

\paragraph{Normal loss} The normal loss $E_n$ is derived from \cref{eq:normal}, penalizing the deviation of $\nabla f_i$ at the sample points from their ground-truth  normals.
\begin{equation}
  E_n := \frac{1}{N} \sum_{i=1}^n \sum_{\mx \in \ms_i}  \|\nabla_\mx f_i(\mx;\theta) - \mn(\mx)\|.
  \label{eq:normal_loss}
\end{equation}

\paragraph{Eikonal equation loss} We introduce the eikonal equation to approximate the signed distance field of $\mS$.
The loss $E_{eik}$~\cite{Gropp2020} encourages $h$ to satisfy the eikonal equation almost everywhere.
\begin{equation}
  E_{eik} := \mathbb{E}_\mx(\|\nabla_\mx h(\mx, \theta)\| - 1)^2.
  \label{eq:eikonal_loss}
\end{equation}

\paragraph{Off-surface loss} $E_o$ is defined to penalize off-surface points whose function values are close to $0$ \cite{sitzmann2019siren}.
\begin{equation}
  E_o := \mathbb{E}_\mx\bigl(\exp(-\alpha |h(\mx;\theta)|)\bigr), \; \alpha \gg 1.
  \label{eq:off_loss}
\end{equation}

\paragraph{Consistency loss} To ensure that $f_i(\mx;\theta)$ is activated in the Boolean tree when evaluating $h (\mx; \theta)$ on $\ms_i$, we introduce the following term to improve this consistency.
\begin{equation}
  E_{cons} :=  \frac{1}{N} \sum_{i=1}^n \sum_{\mx \in \ms_i} |f_i(\mx;\theta) - h (\mx;\theta)|.
  \label{eq:consistency_loss}
\end{equation}

\paragraph{Correction loss}
During training, some sample points may still severely disobey the consistency loss. 
We define a correction loss $E_c$ to suppress this inconsistency with a high penalty.
\begin{equation}
  E_c := \frac{1}{|\mathcal{D}|}\sum_i \sum_{\mx \in \ms_i \cap \mathcal{D}} \beta |h(\mx;\theta) - f_i(\mx;\theta)|, \;  \beta \gg 1.
  \label{eq:correct_loss}
\end{equation}
Here, $\mathcal{D}$ is the point set in which any point violates the constraint: $|f_i(\mx;\theta) - h(\mx;\theta)| < 10^{-5}$.

The total loss is the sum of the above loss terms, and we set $\alpha = \beta = 100$ empirically. We found that there is no need to add the constraints of \cref{eq:hexterior,eq:hinterior} as loss terms to avoid spurious zero isosurfaces when the off-surface loss and geometric initialization~\cite{Atzmon2020} are used. In this way, we can avoid sampling
the ground-truth occupancy field for training. This strategy makes our conversion algorithm friendly to the B-Rep models with imperfections such as unwanted seams caused by exchanging B-Rep data between different software, and extendable to segmented point clouds where accurate interior and exterior information are not available (see \cref{subsec:robustness} and \cref{sec:app}).

\paragraph{Patch grouping} A B-Rep solid may contain many but small surface patches, which results in a large number of Boolean tree leaf nodes and implicit functions. We run a standard graph coloring algorithm on the patch graph of the decomposed patch set to group some disjointed patches and set their corresponding neural functions to be the same. Due to the universal approximation ability of MLP, this strategy is practically useful without compromising conversion quality. To limit the complexity of the group geometry, we set the maximum patch number in a group as $6$. This grouping strategy reduces the number of implicit functions to around $50\%$ on average in our experiments and results in a smaller network (fewer neurons in the last layer) and a shorter training time (10\% faster).

%% file: figures/network.tex
\begin{figure}[t]
    \centering
    \begin{overpic}[width=0.8\columnwidth]{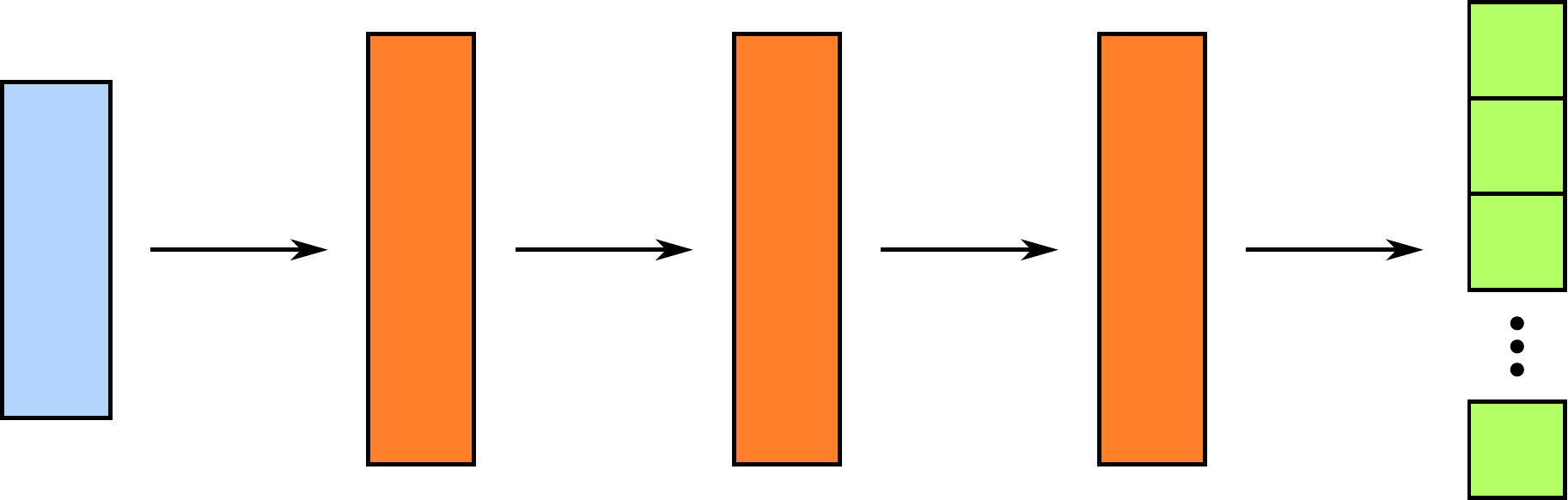}
        \put(2.5,30){\small \textbf{3}}
        \put(3,2){\small $\mx$}
        \put(23.5,32){\small \textbf{256}}
        \put(23.5,-1.5){\small {MLP}}
        \put(47,32){\small \textbf{256}}
        \put(46.5,-1.5){\small {MLP}}
        \put(70,32){\small \textbf{256}}
        \put(69.8,-1.5){\small {MLP}}
        \put(100.5,28){\small \textbf{$f_1$}}
        \put(100.5,22){\small \textbf{$f_2$}}
        \put(100.5,15){\small \textbf{$f_3$}}
        \put(100.5,2){\small \textbf{$f_L$}}
        \put(9,19){\small {SoftPlus}}
        \put(32,19){\small {SoftPlus}}
        \put(55,19){\small {SoftPlus}}
    \end{overpic} 
    \caption{Our network architecture.
    }
    \Description{Our network architecture.}
    \label{fig:network} \vspace{-4mm}
\end{figure}

%% file: src/results.tex
\section{Experiments and Analysis} \label{sec:result}
We conducted a series of experiments and ablation studies to evaluate the efficacy and robustness of our approach and its superiority over other alternative methods.

\subsection{Experiment setup} \label{subsec:setup}

\input{figures/abc}
\paragraph{Dataset} We choose the ABC dataset~\cite{ABC} as our testbed dataset, in which each B-Rep model is a collection of parameterized curve segments and surface patches, including NURBS patches. We choose the first chunk of the ABC dataset that contains \num[group-separator = {,}]{10000} B-Rep models as our testbed. Some models are illustrated in \cref{fig:abc}. We filter out non-manifold and non-closed models, and very simple models like boxes and cylinders. For models with multiple disconnected components, we treat each component as a model instance for conversion, and the resulting NH-Reps can be easily combined via union. The total number of model instances is \num[group-separator = {,}]{10935}. \SI{10}{\percent} models contain NURBS patches. Among them, \num{24} models require patch decomposition during tree construction. We utilized the discretized mesh provided in the ABC dataset for patch decomposition, where the mesh contains u-v coordinates of the parametric patch. The average and maximal surface patch numbers for a model instance are \num{24.07} and 199, respectively. They are reduced to \num{10.69} and 144 by our patch grouping algorithm. \cref{fig:3dtree} illustrate the Boolean trees of two models computed by our method.
\input{figures/3dtree}

\paragraph{Training data preparation}
We randomly sample \myint{50000} points uniformly from each B-Rep model for training, and the corresponding surface normals are assigned to the sampled points. The set of these sample points is denoted as $PS$. On average, we sample $\ceil{50000/L}$ points on each patch, where $L$ is the number of patches in the decomposed patch set. We also ensure that each patch contains at least 50 sampled points, to avoid extremely insufficient sampling. The sampled point cloud is normalized to fit inside a $[-0.9,0.9]^3$ box.

\paragraph{Training details} We train the network to convert a single B-Rep solid to NH-Rep, on an Nvidia GTX 1080 Ti GPU with the PyTorch framework \cite{paszke2017automatic}. We use Adam Optimizer for \num{15000} iterations with an initial learning rate of \num{0.005} scheduled to drop by a factor of 2 every \myint{2000} iterations. The correction loss term is enabled after \myint{10000} iterations. We randomly select \myint{16384} points from $PS$ in each iteration to calculate $E_p$, $E_n$, $E_o$, and $E_{cons}$, where $16384/L$ points are sampled on each patch. To calculate $E_{eik}$ and $E_o$, we randomly sample a set of points within $[-1,1]^3$ in each iteration. These points fall into two categories: \emph{local samples} and \emph{global samples}. Local samples are points perturbed from the sampled points with a normal distribution (mean: 0, stdev: $\sigma$) along a random direction, where $\sigma$ is the shortest distance from the given sample point to the densely sampled point cloud. In total, the local samples contain \myint{16384} points. Global samples contain \myint{2048} points randomly sampled in $[-1,1]^3$ according to a normal distribution (mean: \num{0}, stdev: \num{1.8}).
$E_{eik}$ uses all local and global sample points, and $E_o$ uses global samples only.  The average training time for a single model is about 10 minutes.

\paragraph{Isosurface extraction}
For evaluating the approximation quality of NH-Rep to the input B-Rep boundary, we extracted the isosurface of NH-Rep as a polygon mesh with feature preservation by the advanced isosurfacing algorithms~\cite{Kobbelt2001,Ju2002} as follows. On each edge of the cube, we compute its intersection point with $h=0$, and also record the functional gradient as the point normal. All intersection computations are performed in parallel in GPU. The intersection points with normals are fed to the feature-aware marching cube algorithm to extract a mesh surface. The default grid resolution is $256^3$ and we increase the resolution to $512^3$ automatically if the isosurfacing algorithm generates very long mesh edges, indicating that the model may contain extremely narrow geometry. For models with very sharp features, we found that the dual contour algorithm \cite{Ju2002} is more stable and accurate than the algorithm of \cite{Kobbelt2001}. In our implementation, we used an octree structure to speed up isosurface extraction.
Here, we note that recent learning-based isosurfacing works \cite{chen2021neural,chen2022neural}
are different from our work. These methods take discrete signals as input without access to surface gradients, such as a distance field grid, and predict mesh vertex location and edge crossing in cube cells for robustly reconstructing mesh facets and sharp edges. They are designed for mesh reconstruction and do not have access to ground-truth isosurfaces during the test phase. In our work, any isosurface of NH-Rep can be accurately evaluated, and thus the isosurface extraction step is just a conversion from implicit surfaces to mesh format.

\paragraph{Evaluation metrics}
We measure the quality of implicit conversion with respect to the following metrics.
\begin{itemize}[leftmargin=*,nosep]\setlength\itemsep{1mm}
    \item[-] Chamfer distance (\textbf{CD}) and two-side Hausdorff Distance (\textbf{HD}) between the extracted zero surface and the corresponding B-Rep surface.  \myint{50000} points are randomly sampled on both surfaces for computation.
    \item[-] Average normal error (\textbf{NAE}): Chamfer distance with respect to surface normals, using the same sample points as used for \textbf{CD}.
    \item[-] Feature Chamfer distance (\textbf{FCD}) and feature angle error (\textbf{FAE}). We first detect sharp feature edges on $\mathbf{M}$ with the dirhedral angle threshold: $\delta_s=\ang{30}$, then uniformly sample points on the detected feature edges where the point distance interval is $0.004$. We also record the dihedral angle for each edge sample point. A similar sampling is also done on $\partial\mathcal{S}$. The Chamfer distance between these sample points defines \textbf{FCD}, and the sum of the absolution error of the dihedral angles between any nearest paired points in the \textbf{FCD} computation defines \textbf{FAE}.
    \item[-] Occupancy IoU (\textbf{IoU}) with respect to the \emph{ground-truth} mesh model provided in the ABC data set, calculated on a set of random sample points in $[-1,1]^3$.
    \item[-] The sum of the relative error of the predicted SDF with the real SDF (\textbf{DE}), calculated on a set of random sample points in $[-1,1]^3$. The real SDF is calculated on the \emph{ground-truth} mesh model.
\end{itemize}
Here, \textbf{CD} and \textbf{HD} measure the overall quality of the zero surfaces; \textbf{FCD} and \textbf{FAE} measure the sharp feature quality of the zero surfaces; \textbf{DE} and \textbf{IoU} measure the implicit field quality. The exact metric formulas are provided in \cref{appendix:metric}.

\input{figures/benchmark}

\subsection{Experiment analysis} \label{subsec:analysis}

\paragraph{Benchmark and comparison} We applied our method to convert \myint{10935} B-Rep models to implicit representations. For comparison, we choose two classic feature-aware reconstruction methods: Screened Poisson Reconstruction (SPR) \cite{Kazhdan2013} and Robust Implicit Moving Least Squares (RIMLS) \cite{Oztireli2009}, and three representative learning-based implicit approaches: IGR~\cite{Gropp2020}, SIREN~\cite{sitzmann2019siren}, Neural Spline (NS)~\cite{williams2021neural}. The input to these competitive methods is the same sample points from the B-Rep model as we described in \cref{subsec:setup}.
Note that these methods cannot leverage patch adjacency information as ours.
The SPR and RIMLS implementations are from MeshLab~\cite{meshlab} with default settings. For SPR, its interpolation weight is set to \num{50} for better feature preservation. For IGR, SIREN, and NS, we use their default network and the suggested optimal parameter settings. The average training time for a single model is 40 minutes for IGR, 20 minutes for SIREN, and 1 minute for NS. Our network parameter size is 137k on average, which is smaller than SIREN (198k), IGR (1837k) and NS (250k). For IGR and SIREN, we apply our surface extraction algorithm to recover sharp features, as their implicit functions are easy to access. For SPR, RIMLS, and NS, we use the marching cube algorithm supplied by their original implementation.  For SIREN, we also remove unwanted small components and pick the largest component for evaluation.  The failure cases of RIMLS (126 models) and SPR (3 models) are not counted in the performance report.
\input{figures/morepatch}
\input{tables/abcbenchmark_new}

The performance of all methods is reported in \cref{tab:abcbenchmark}. SPR, IGR, and our approach have comparable quality in terms of CD, HD and NAE, while RIMLS, SIREN, and NS perform worse. Our approach achieves the best performance on the sharp-feature-related metrics (FCD, FAE) and IoU. The DE of ours is worse than IGR, as NH-Rep does not recover the exact signed distance function due to the use of Boolean operations. The visual comparison shown in \cref{fig:benchmark} further confirms the superiority of our method in preserving sharp features and recovering surface geometry. The robustness of our conversion method to B-Rep inputs with many surface patches is also illustrated in \cref{fig:benchmark}-bottom and \cref{fig:morepatch} where the inputs have more than 100 patches. In our supplemental material, we provide metric histograms and more conversion results for further analysis.

\input{tables/equaltime}
\input{figures/fig_equaltime}

\paragraph{Equal-time comparison with SPR and NS}
As both NS and SPR with their default settings have less running time than our approach, we feed more sample points (5M points) and enable longer iterations (SPR: 800 iterations, NS: 200 iterations) and higher resolution for mesh extraction (NS: depth = 9, SPR: depth = 10), to maximize their conversion capacity.  We used the complex model shown at the bottom of \cref{fig:benchmark} as a test example. NS and SPR took 40 and 10 minutes to compute, respectively. The results of the comparison are reported in \cref{tab:equaltime} and \cref{fig:equaltime}. NS produces bumpy geometry and spurious zero surface, and SPR achieves comparable accuracy (CD\&HD) to our result trained with 50K points, but still has larger feature errors and contains tiny fluctuation in the flat region; furthermore, its implicit field storage is huge, exceeding \SI{1.8}{GB} memory.

\input{tables/feature}

\paragraph{Training with sharp feature samples}
Our default point sampling strategy does not include many sharp feature points for training. We design an additional test to check whether IGR and SIREN training with more sharp feature points can outperform our method. We selected 100 B-Rep models randomly from our benchmark dataset and added 10k points sampled at feature edges for each model, with the original sampled points. As seen in the qualitative evaluation (\cref{tab:feature}), IGR and SIREN with these additional inputs cannot yet compete with our method which does not utilize sharp feature points, while these additional inputs slightly improve our method.

\paragraph{Comparison with CSG-based methods} We also compared our method with BSP-Net~\cite{bspnet}, CSG-Stump~\cite{ren2021csg}, and CAPRI-Net~\cite{yu2021capri}, which predict the CSG tree and a set of primitives to approximate the input solid.
As these methods were trained on ShapeNet~\cite{shapenet2015}, we selected five typical models from their test sets and used their trained networks and prepared input data for comparison.
The inputs are voxel cells (resolution $64^3$) for BSP-Net, \num[group-separator = {,}]{2048} points for CSG-Stump, and \num[group-separator = {,}]{8192} points with normal vectors for CAPRI-Net.
For our method, we detect sharp features of the input mesh models and segment each mesh into a set of patches to construct a B-Rep for implicit conversion. We extracted the zero isosurfaces of these methods for visual comparison. As seen in \cref{fig:bspnet}, BSP-Net, CSG-Stump and CAPRI-Net approximate the inputs roughly with compact and simple elements but cannot accurately represent the input solids; while our method achieves more accurate implicit conversion results.

\input{figures/bspnet}

\subsection{Ablation study}
We validate our loss design and network design through a set of ablation studies, performed on 100 B-Rep models randomly selected from the benchmark dataset.

\paragraph{Efficacy of loss terms}
We validate the efficacy of each loss term by dropping one of them during training. The performance of these ablations is reported in \cref{tab:loss}. We have the following observations.
\begin{itemize}[leftmargin=*,nosep]\setlength\itemsep{1mm}
    \item The missing of $E_n$ produces the highest errors in all metrics, since the normal orientation of $h$ on the surface points could be very different from the ground truth. \cref{fig:ablation}-(a) illustrates typical artifacts such as distorted and non-smooth geometry, even additional parts, and missing sharp features.
    \item Without $E_o$, the network has less ability to constrain $h=0$ on the input patches; thus, it may produce additional components as shown in \cref{fig:ablation}-(b), and lead to a higher HD error.
    \item Without $E_{cons}$ and $E_c$, $f_i$ may be inactivated when evaluating $h$ on patch $\ms_i$, resulting in a higher HD error than our default setting, as shown in \cref{fig:ablation}-(c).
    \item Our default setting achieves the lowest error in all metrics, being more faithful to the input.
\end{itemize}

\input{tables/ablation_new100}

\input{figures/ablation}

\paragraph{Impact of network size}
We tested our network performance with different numbers of MLP layers and neurons. The network configuration is indicated as $N \times M$, $N$ is the number of neurons in each MLP layer, and $M$ is the number of MLP layers. \cref{tab:networksize} reports the average performance.
With a fixed $M$ ($M=3$), a larger $N$ helps to improve all metrics; when $M=7$, the improvement brought by a larger $N$ is less significant and worse on some metrics such as HD and FAE possibly due to overfitting.
Our default setting $256\times 3$ balances network performance and network size. It obtains the best performance on feature preservation and is slightly worse than the configuration of $256\times 7$ in other metrics.

\input{tables/networksize_new}

\paragraph{Separated MLPs for individual patches}
In our network architecture, the MLP layers are shared to predict individual $f_i$s. We conducted an ablation study on two models (see \cref{fig:ablation_split})  to
verify whether the use of separated MLPs for individual patches could improve the accuracy of implicit conversion. We find that the implicit conversion qualities of both versions are comparable (see \cref{tab:fig11_12}).
However, the network size of using separated MLPs is much larger, as it is proportional to the number of patches. Our shared MLP strategy provides a great balance between model size and implicit conversion quality.

\input{figures/ablation_split}

\input{tables/fig11_12}

\subsection{Robustness test} \label{subsec:robustness}

\paragraph{Data noise}
To test the robustness of our algorithm to noise, we add random noise in the range of $[-0.018, 0.018]$ to point positions along normal directions and perturb normal directions within the angle interval $[\ang{-3}, \ang{3}]$. Here, for our method, we assume that the patch information and feature convexity are correct, and we also drop the correction loss in our training because this loss function is sensitive to noise. \cref{fig:noise} shows the reconstruction results generated by our method, as well as SIREN and IGR, on two noisy inputs.  The results show that our approach and IGR are less affected by noise, whereas our approach achieves the best result possibly because we have fewer MLP layers than IGR that avoid the overfitting problem, and our CSG operations are good for recovering sharp features.
\input{figures/noise}

\paragraph{Segmented point clouds}
In practice, CAD models are obtained by scanning, in point cloud format, without B-Rep information. It is possible to segment the point cloud first, then use our method to convert it to NH-Rep. In \cref{fig:ourseg}, we utilize ParseNet~\cite{sharma2020parsenet} to obtain the segmentation and recognize the type of feature curves (convexity or concavity) between two adjacent patches according to the majority of normal variations at the boundary points. The input models are selected from the ParseNet test set. The zero isosurfaces in the figure show that our method recovers the CAD mesh in good quality if the segmentation is reliable. For the imperfect segmentation shown in the bottom row, our approach still approximates the input but fails to model some sharp features due to the wrong segmentation (dark green region), which consists of two disjoint parts that cross the feature region.

\input{figures/ourseg}

\input{figures/patchdecom}
\input{tables/decomtab}

\paragraph{Sensitivity to patch decomposition}
In our patch decomposition algorithm, random patch selection can introduce different decomposed patch sets, but has a minor impact on the quality of converted NH-Reps. We select a complicated model (\cref{fig:repair3d}-a) that requires patch composition to evaluate the impact of different decompositions. Two different decomposed patch sets are generated (see \cref{fig:repair3d}-b top and bottom).
Both the zero isosurfaces of the resulting NH-Reps (\cref{fig:repair3d}-c) are close to ground truth, and the quality of two different NH-Reps is comparable (see \cref{tab:repair3d}).

\input{figures/scalability}

\paragraph{Scalability}
For CAD models with a large number of patches in the ABC dataset, we find that many patches are very small and narrow, and many of the shared feature curves are not sharp. If we naively apply our implicit conversion algorithm, dense sample points on and around small patches are needed to keep the approximation error low.  A practical way is to merge small patches with its neighbors to reduce the patch number if their shared edges are \emph{smooth}. We tested this strategy on a model with 868 patches. \cref{fig:scalability} shows that the total number of patches can be reduced to 11, and our conversion algorithm can perform well.

%% file: figures/abc.tex
\begin{figure}[t]
    \centering
    \begin{overpic}[width=0.95\columnwidth]{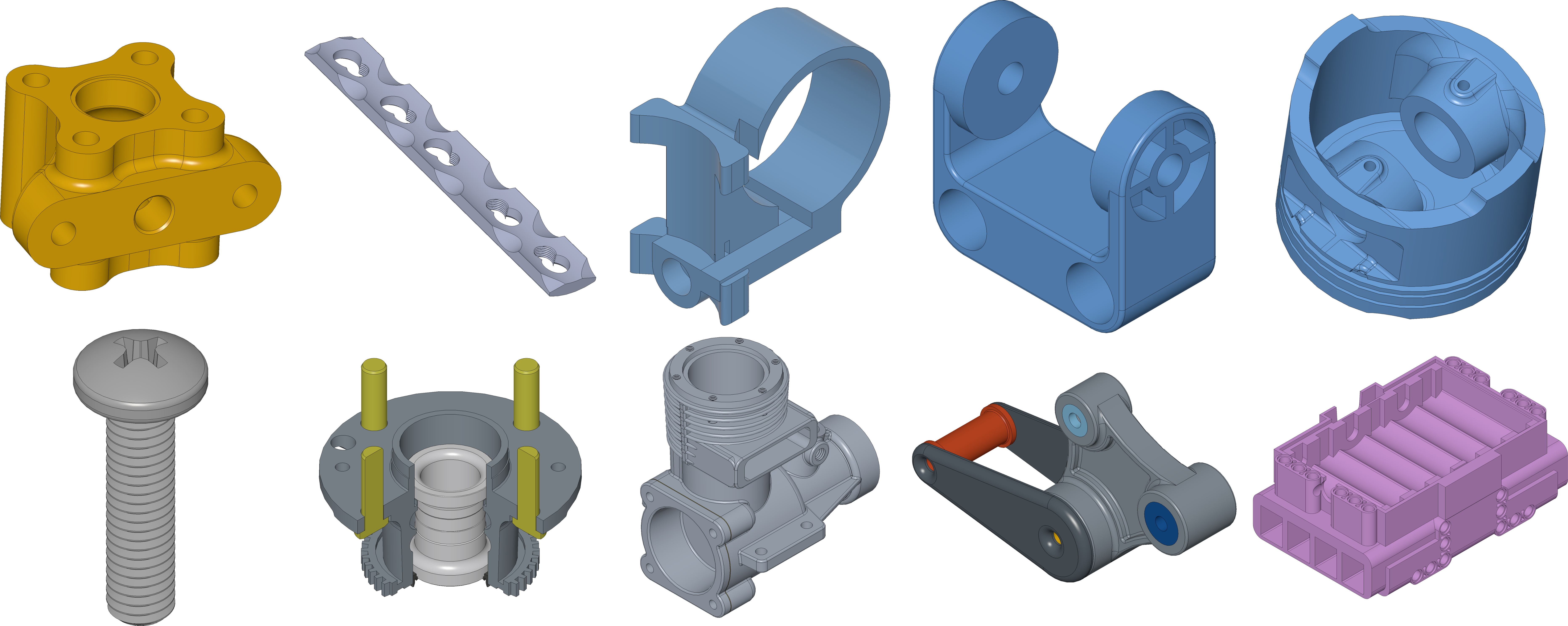}
    \end{overpic} 
    \caption{B-Rep solid samples in the ABC dataset~\cite{ABC}.}
    \label{fig:abc} \vspace{-2mm}
    \Description{B-Rep solid samples in the ABC dataset}
\end{figure}

%% file: figures/3dtree.tex
\begin{figure*}[t]
    \centering
    \begin{overpic}[width=\linewidth]{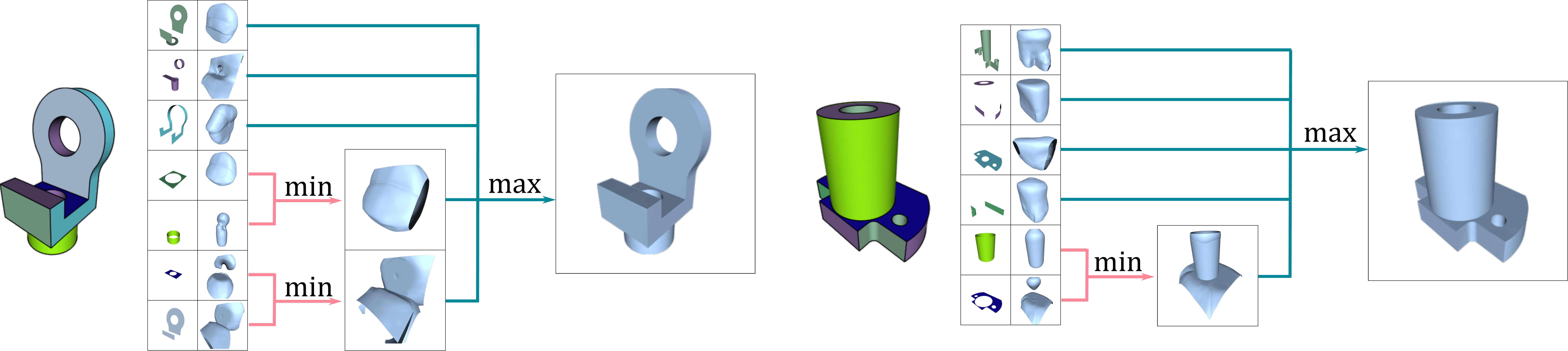}
        \put(10.5,-1){\small $s_i$}
        \put(62.9,0.7){\small $s_i$}
    \end{overpic} 
    \caption{Boolean trees of two B-Rep solids.  B-Rep patches are rendered in different colors. The zero isosurfaces of the Boolean functions at tree nodes are also illustrated (truncated by the bounding box).}
    \Description{Boolean trees of two B-Rep solids.}
    \label{fig:3dtree} \vspace{-2mm}
\end{figure*}

%% file: figures/benchmark.tex
\begin{figure*}[t]
    \centering
    \begin{overpic}[width=0.95\linewidth]{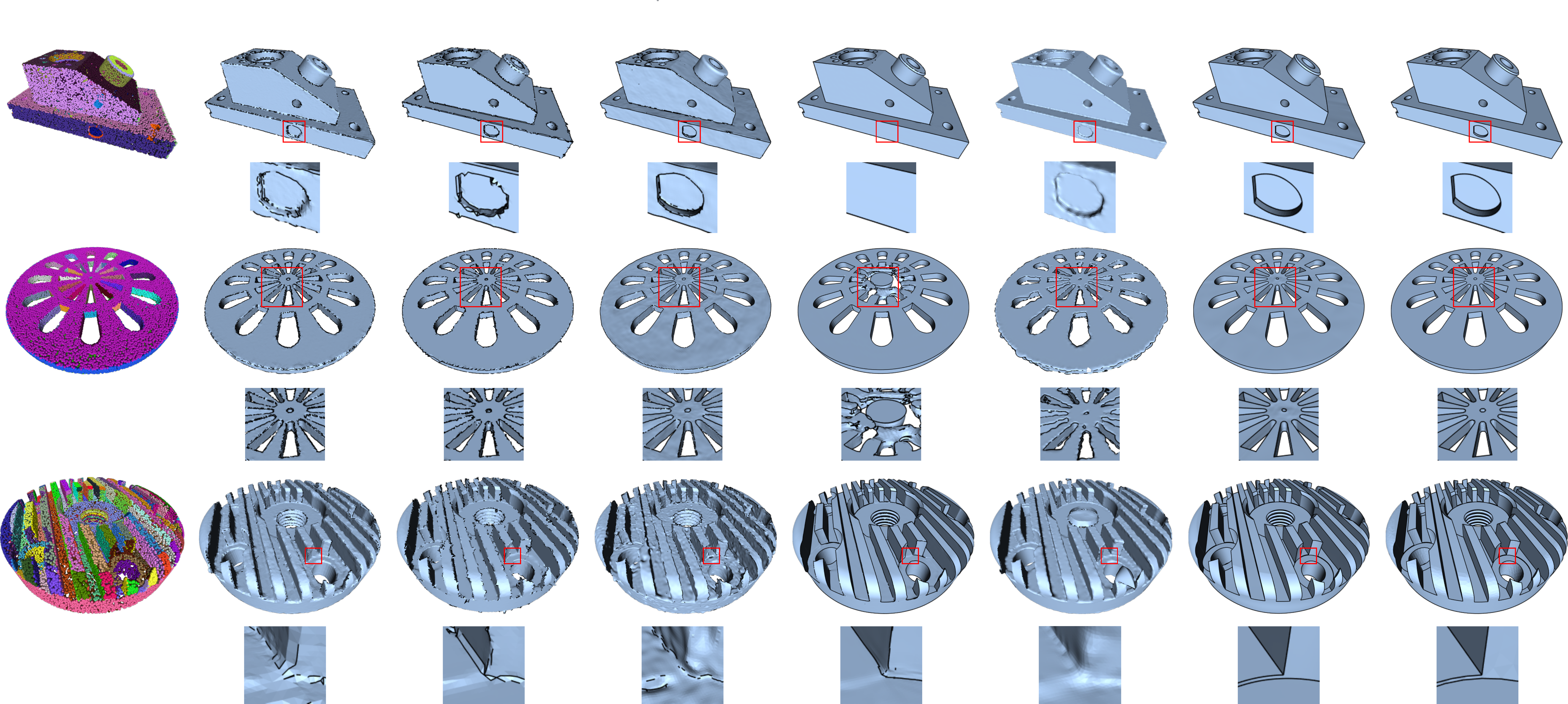}
        \put(3,-2.5){\small {\textbf{Input}}}
        \put(16.5,-2.5){\small {\textbf{SPR}}}
        \put(28,-2.5){\small {\textbf{RIMLS}}}
        \put(41,-2.5){\small {\textbf{SIREN}}}
        \put(54.5,-2.5){\small {\textbf{IGR}}}
        \put(64,-2.5){\small {\textbf{Neural Spline}}}
        \put(79.5,-2.5){\small {\textbf{Ours}}}
        \put(89,-2.5){\small {\textbf{Ground truth}}}

        \put(0,15.5){\small {\textbf{\#patches: 184}}}
        \put(0,30){\small {\textbf{\#patches: 52}}}
        \put(0,43){\small {\textbf{\#patches: 65}}}
    \end{overpic}
    \vspace{1mm}
    \caption{Visual comparison of different approaches for implicit conversion. We render the input B-Rep models as point clouds that are the inputs to the compared methods, different B-Rep patches in random colors. The detected feature edges are illustrated in black. Only our method can reproduce correct sharp features after implicit conversion. The zoom-in views highlight the artifacts of the results from other methods.}
    \label{fig:benchmark}  \vspace{-2mm}
\end{figure*}

%% file: figures/morepatch.tex
\begin{figure}[t]
    \centering
    \begin{overpic}[width=0.95\columnwidth]{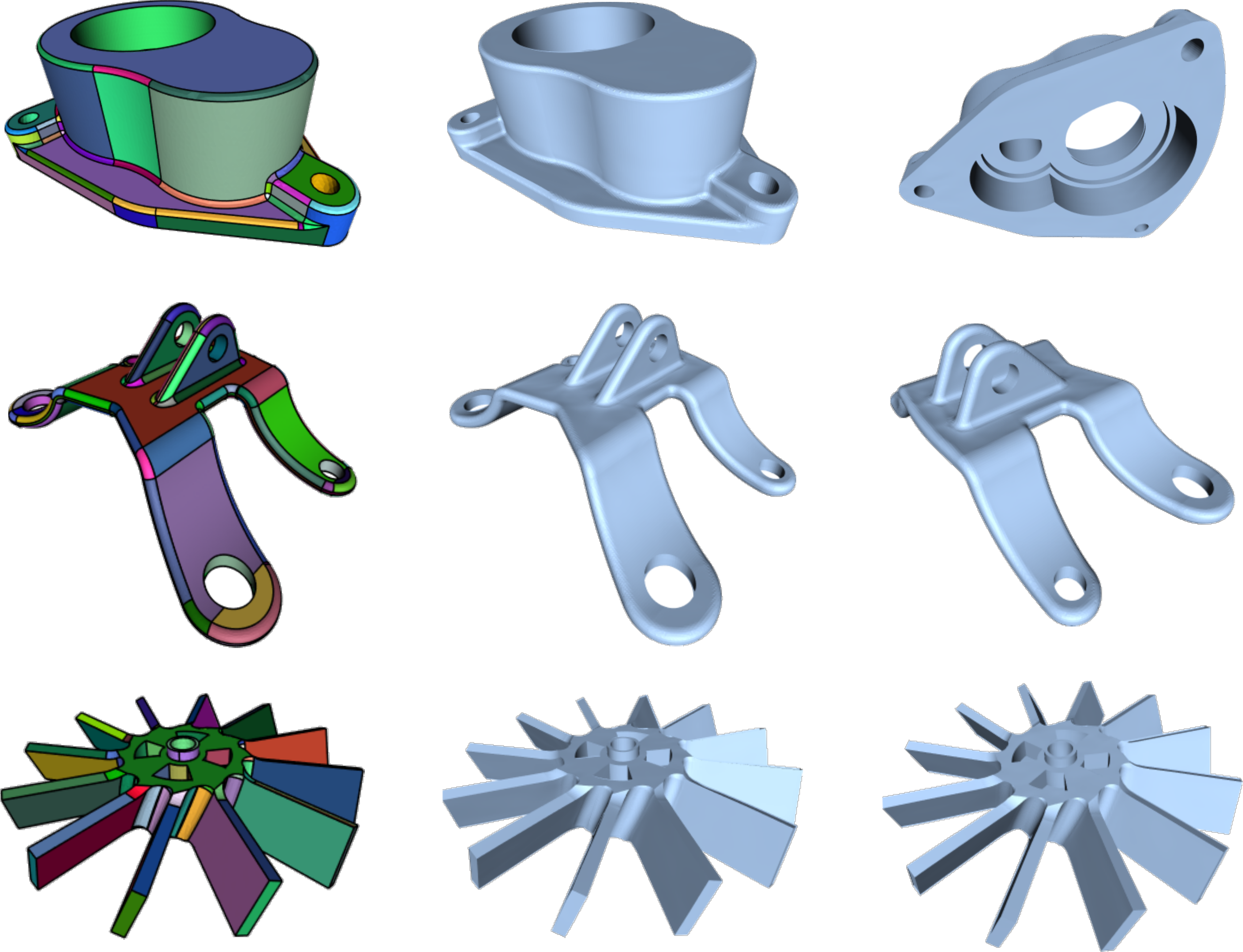}
        \put(10, -2.3){\small input}
        \put(48,-2.3){\small two rendered views of $h(\mx)=0$}
    \end{overpic} 
    \caption{NH-Rep conversion of three complicated B-Rep models. The patch numbers (from top to bottom) are: 121, 178, 191. The zero isosurfaces of $h(\mx)$ are rendered from two different views for better visualization.}
    \label{fig:morepatch} \vspace{-4mm}
\end{figure}

%% file: tables/ABCbenchmark_new.tex
\begin{table}[t]
    \centering
    \caption{Quantitative evaluation of implicit conversion of different methods on \myint{10935} B-Rep models.  CD, HD and FCD are scaled by $10^3$.
        \label{tab:abcbenchmark}}
    \scalebox{0.85}{
        \begin{tabular}{c|rrrrrrr}
            \toprule
            \thead{Method} & \thead{CD   $\downarrow$} & \thead{HD  $\downarrow$} & \thead{NAE $\downarrow$} & \thead{FCD $\downarrow$} & \thead{FAE $\downarrow$} & \thead{DE $\downarrow$} & \thead{IoU $\uparrow$} \\
            \midrule
            SPR            & \mybold{5.07}             & \mybold{21.5}            & \ang{4.14}               & 14.5                     & \ang{43.3}               & n/a                     & \mybold{0.979}         \\
            RIMLS          & 6.17                      & 32.6                     & \ang{4.07}               & 20.7                     & \ang{31.1}               & n/a                     & 0.325                  \\
            SIREN          & 14.3                      & 87.3                     & \ang{4.57}               & 25.6                     & \ang{28.4}               & 0.657                   & 0.889                  \\
            IGR            & 7.10                      & 41.0                     & \ang{2.92}               & 13.5                     & \ang{9.61}               & \mybold{0.039}          & 0.965                  \\
            NS             & 16.7                      & 95.7                     & \ang{6.97}               & 30.3                     & \ang{48.4}               & n/a                     & 0.762                  \\
            ours           & 5.31                      & 24.3                     & \myboldangle{2.42}       & \mybold{7.07}            & \myboldangle{3.69}       & 0.064                   & \mybold{0.979}         \\
            \bottomrule
        \end{tabular}
    }
\end{table}

%% file: tables/equaltime.tex
\begin{table}[t]
    \centering
    \caption{Quantitative evaluation of implicit conversion of different methods on \cref{fig:benchmark}-bottom model under equal-time setting.}
    \label{tab:equaltime}
    \scalebox{0.85}{
        \begin{tabular}{c|rrrrrrr}
            \toprule
            \thead{Method} & \thead{CD   $\downarrow$} & \thead{HD  $\downarrow$} & \thead{NAE $\downarrow$} & \thead{FCD $\downarrow$} & \thead{FAE $\downarrow$} & \thead{DE $\downarrow$} & \thead{IoU $\uparrow$} \\
            \midrule
            SPR            & \mybold{7.86}             & \mybold{28.2}            & \myboldangle{6.00}       & 2.28                     & \ang{42.9}               & n/a                     & \mybold{0.997}         \\
            NS             & 8.89                      & 154                      & \ang{7.45}               & 5.50                     & \ang{45.3}               & n/a                     & 0.326                  \\
            ours           & 7.93                      & 30.6                     & \ang{6.30}               & \mybold{2.07}            & \myboldangle{7.59}       & 0.0668                  & 0.987                  \\
            \bottomrule
        \end{tabular}
    }
\end{table}

%% file: figures/fig_equaltime.tex
\begin{figure}[t]
    \centering
    \begin{overpic}[width=0.85\columnwidth]{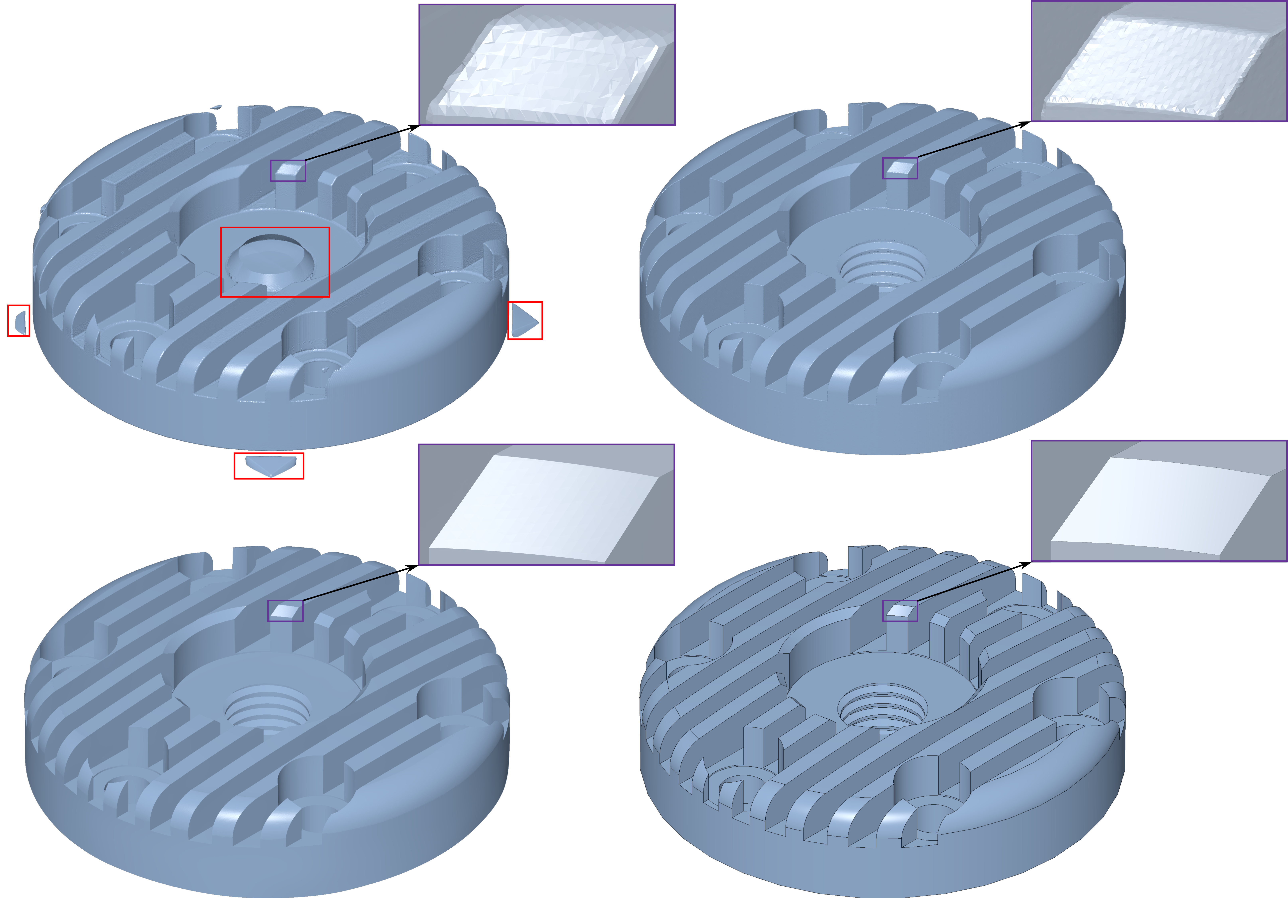}
        \put(19,65){\small NS}
        \put(17,-1){\small Ours}
        \put(67.5,65){\small SPR}
        \put(60,-1){\small Ground truth}
    \end{overpic}
    \caption{Visual comparison of different approaches for implicit conversion under equal-time setting. Our method reproduces features more accurately than other methods.}
    \label{fig:equaltime} \vspace{-3mm}
\end{figure}

%% file: tables/feature.tex
\begin{table}[t]
    \caption{Qualitative evaluation of implicit conversion of different methods using additional sharp feature points on 100 B-Rep models. \textbf{Fea.} indicates whether these additional points are used.}.
    \label{tab:feature}
    \centering
    \scalebox{0.8}{
        \begin{tabular}{ccrrrrrrr}
            \toprule
            \thead{Method} & \thead{Fea.} & \thead{CD  $\downarrow$} & \thead{HD  $\downarrow$} & \thead{NAE $\downarrow$} & \thead{FCD  $\downarrow$} & \thead{FAE $\downarrow$} & \thead{DE $\downarrow$} & \thead{IoU $\uparrow$} \\
            \midrule
            IGR            & \checkmark   & 7.21                     & 42.6                     & \ang{3.11}               & 5.16                      & \ang{9.94}               & 0.0444                  & 0.981                  \\
            SIREN          & \checkmark   & 7.07                     & 29.5                     & \ang{4.17}               & 8.49                      & \ang{29.4}               & 0.4570                  & 0.968                  \\
            ours           & $\times$     & 5.49                     & 22.7                     & \ang{2.67}               & 3.80                      & \ang{3.75}               & 0.0583                  & 0.989                  \\
            ours           & \checkmark   & \mybold{5.42}            & \mybold{21.7}            & \myboldangle{2.57}       & \mybold{3.45}             & \myboldangle{3.43}       & \mybold{0.0526}         & \mybold{0.992}         \\
            \bottomrule
        \end{tabular}
    }
    \vspace{-4mm}
\end{table}

%% file: figures/bspnet.tex
\begin{figure}[t]
    \centering
    \begin{overpic}[width=\columnwidth]{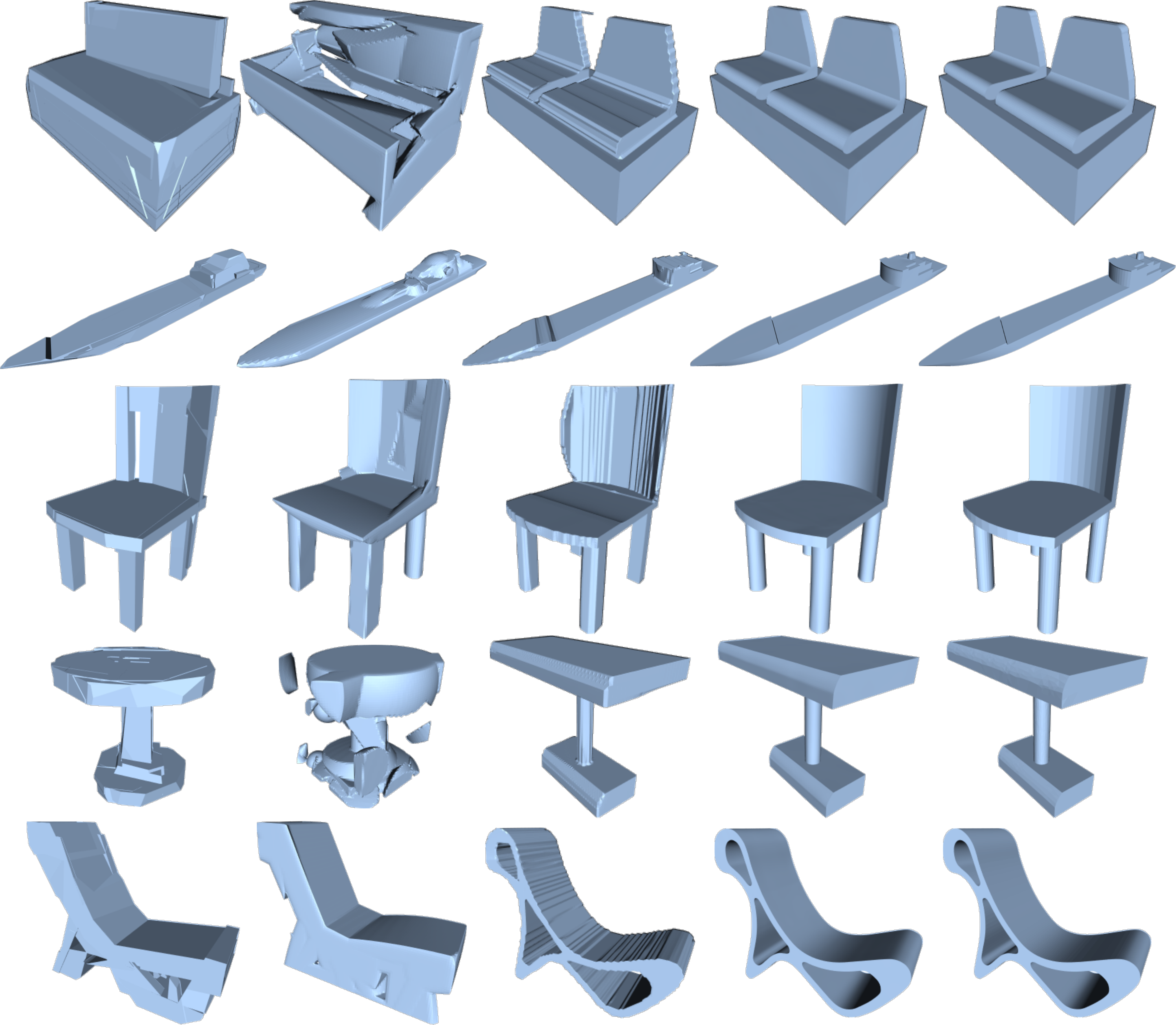}
        \put(4.5,-3.5){\small BSP-Net}
        \put(23,-3.5){\small CSG-Stump}
        \put(43,-3.5){\small CAPRI-Net}
        \put(67,-3.5){\small Ours}
        \put(80,-3.5){\small Ground truth}
    \end{overpic}\vspace{1pt}
    \caption{Comparison with BSP-Net, CSG-Stump and CAPRI-Net on implicit conversion. Inputs are selected from ShapeNet. We illustrate the boundary surfaces extracted from different methods. BSP-Net, CSG-Stump and CAPRI-Net fail to approximate CAD models accurately due to the limited representation power and inaccurate prediction. }
    \label{fig:bspnet} \vspace{-5mm}
\end{figure}

%% file: tables/ablation_new100.tex
\begin{table}[t]
    \centering \caption{Ablation study of loss terms on 100 models.} \label{tab:loss}
    \scalebox{0.8}{
        \begin{tabular}{crrrrrrr}
            \toprule
            \thead{Config.}       & \thead{CD  $\downarrow$} & \thead{HD  $\downarrow$} & \thead{NAE $\downarrow$} & \thead{FCD  $\downarrow$} & \thead{FAE $\downarrow$} & \thead{DE $\downarrow$} & \thead{IoU $\uparrow$} \\
            \midrule
            w/o $E_n$             & 26.0                     & 207                      & \ang{12.7}               & 39.7                      & \ang{24.3}               & 0.3440                  & 0.625                  \\
            w/o $E_o$             & 6.07                     & 41.3                     & \ang{2.75}               & 4.88                      & \ang{4.10}               & 0.0612                  & 0.986                  \\
            w/o $E_c$\&$E_{cons}$ & 5.50                     & 24.7                     & \ang{2.69}               & 3.82                      & \ang{3.85}               & 0.0868                  & \mybold{0.989}         \\
            default               & \mybold{5.49}            & \mybold{22.7}            & \myboldangle{2.67}       & \mybold{3.80}             & \myboldangle{3.75}       & \mybold{0.0583}         & \mybold{0.989}         \\
            \bottomrule
        \end{tabular}
    }
    \vspace{-2mm}
\end{table}

%% file: figures/ablation.tex
\begin{figure*}[t]
    \centering
    \begin{overpic}[width=0.95\textwidth]{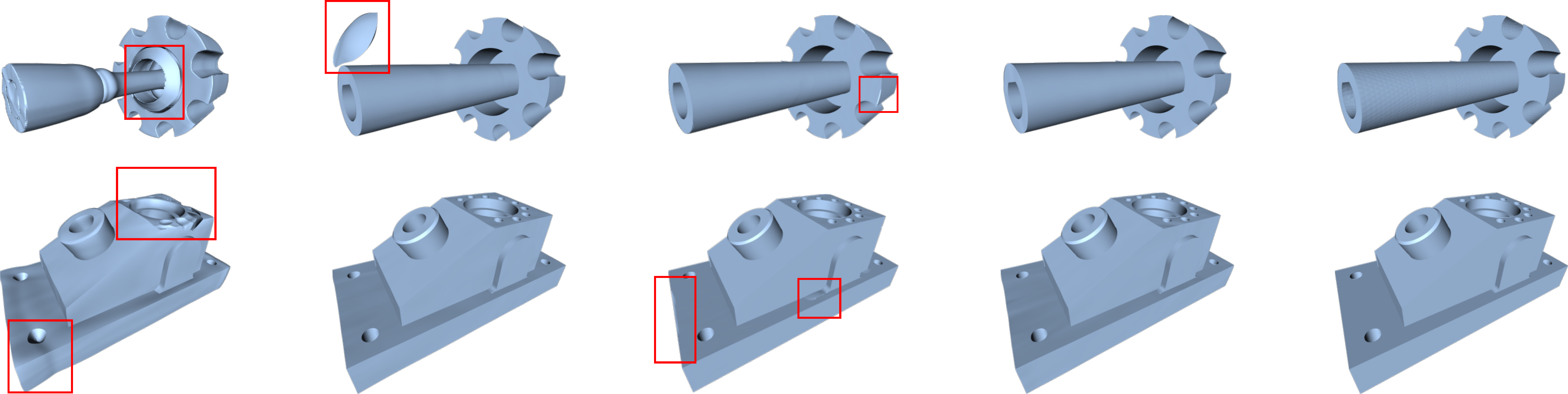}
        \put(3,-1.5){ \small \textbf{(a)}: w/o $E_n$}
        \put(25,-1.5){\small \textbf{(b)}: w/o $E_o$}
        \put(43,-1.5){ \small \textbf{(c)}: w/o $E_{cons}\&E_c$}
        \put(68,-1.5){\small \textbf{(d)} Default}
        \put(89,-1.5){\small Ground truth}
    \end{overpic} 
    \caption{Ablation study of loss terms. Artifacts are highlighted by red boxes.
    }
    \label{fig:ablation} \vspace{-2mm}
    \Description{Ablation study of loss terms}
\end{figure*}

%% file: tables/networksize_new.tex
\begin{table}[t]
    \caption{Network performance with different network configurations on 100 models. $256\times 3$ is our default setting. }
    \label{tab:networksize}
    \centering
    \scalebox{0.85}{
        \begin{tabular}{rrrrrrrr}
            \toprule
            \thead{Config.} & \thead{CD  $\downarrow$} & \thead{HD  $\downarrow$} & \thead{NAE $\downarrow$} & \thead{FCD  $\downarrow$} & \thead{FAE $\downarrow$} & \thead{DE $\downarrow$} & \thead{IoU $\uparrow$} \\
            \midrule
            $64\times 3$    & 5.79                     & 34.7                     & \ang{3.06}               & 4.33                      & \ang{4.28}               & 0.0839                  & 0.981                  \\
            $128\times 3$   & 5.66                     & 27.9                     & \ang{2.81}               & 3.98                      & \ang{3.89}               & 0.0669                  & 0.981                  \\
            $256\times 3$   & 5.49                     & 22.7                     & \ang{2.67}               & 3.80                      & \myboldangle{3.75}       & 0.0583                  & 0.989                  \\
            $256\times 7$   & \mybold{5.44}            & \mybold{21.3}            & \myboldangle{2.63}       & 3.81                      & \ang{5.02}               & 0.0569                  & \mybold{0.990}         \\
            $512\times 7$   & 5.46                     & 23.4                     & \ang{2.70}               & \mybold{3.61}             & \ang{6.18}               & \mybold{0.0477}         & 0.988                  \\
            \bottomrule
        \end{tabular}
    }
    \vspace{-2mm}
\end{table}

%% file: figures/ablation_split.tex
\begin{figure}[t]
    \centering
    \vspace{3mm}
    \begin{overpic}[width=\columnwidth]{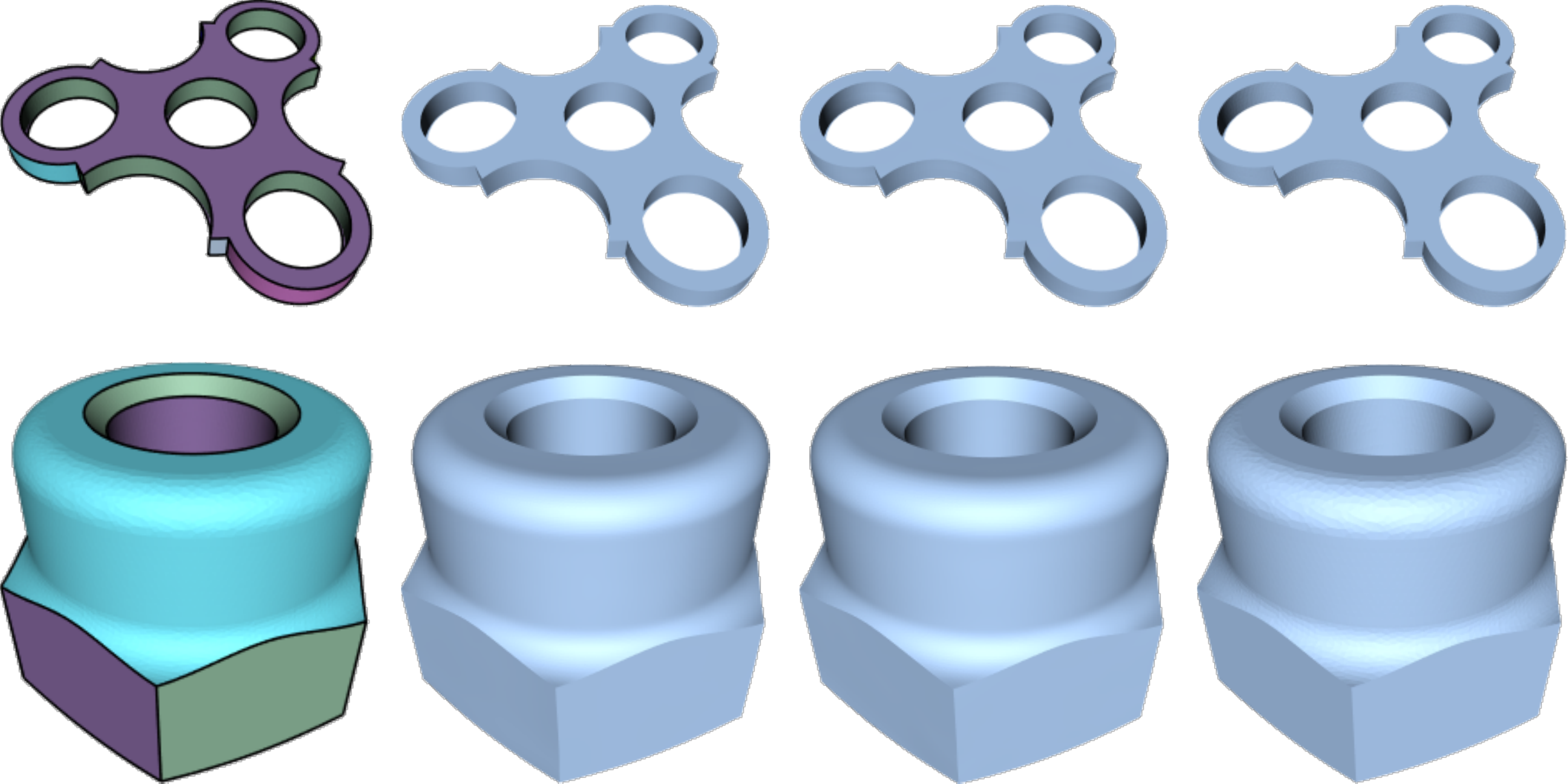}
        \put(7,-3.5){\small Input}
        \put(27,-3.5){\small Separated MLPs}
        \put(55,-3.5){\small Shared MLPs}
        \put(78,-3.5){\small Ground truth}
        \put(0, 2){\small B}
        \put(0, 35){\small A}
    \end{overpic}\vspace{1pt}
    \caption{Ablation study of shared MLPs and separated MLPs on two models.   }
    \label{fig:ablation_split}
    \Description{Ablation study of shared MLPs and separated MLPs}
\end{figure}

%% file: tables/fig11_12.tex
\begin{table}[t]
    \caption{Qualitative evaluation of implicit conversion of models in \cref{fig:ablation_split}. 
    }
    \label{tab:fig11_12}
    \centering
    \scalebox{0.8}{
        \begin{tabular}{ccrrrrrrr}
            \toprule
            \thead{Model} & \thead{Config.} & \thead{CD  $\downarrow$} & \thead{HD  $\downarrow$} & \thead{NAE $\downarrow$} & \thead{FCD  $\downarrow$} & \thead{FAE $\downarrow$} & \thead{DE $\downarrow$} & \thead{IoU $\uparrow$} \\
            \midrule
            A             & sep.            & 4.26                     & 17.4                     & \ang{2.98}               & 1.31                      & \ang{0.522}              & 0.0742                  & 0.989                  \\
            A             & shared          & 4.23                     & 16.3                     & \ang{3.06}               & 1.34                      & \ang{0.625}              & 0.0712                  & 0.990                  \\
            B             & sep.            & 7.90                     & 28.3                     & \ang{1.61}               & 6.97                      & \ang{7.67}               & 0.0590                  & 0.996                  \\
            B             & shared          & 7.86                     & 29.6                     & \ang{1.58}               & 7.22                      & \ang{7.45}               & 0.0303                  & 0.997                  \\
            \bottomrule
        \end{tabular}
    }
    \vspace{-4mm}
\end{table}

%% file: figures/noise.tex
\begin{figure}[t]
    \centering
    \begin{overpic}[width=0.95\columnwidth]{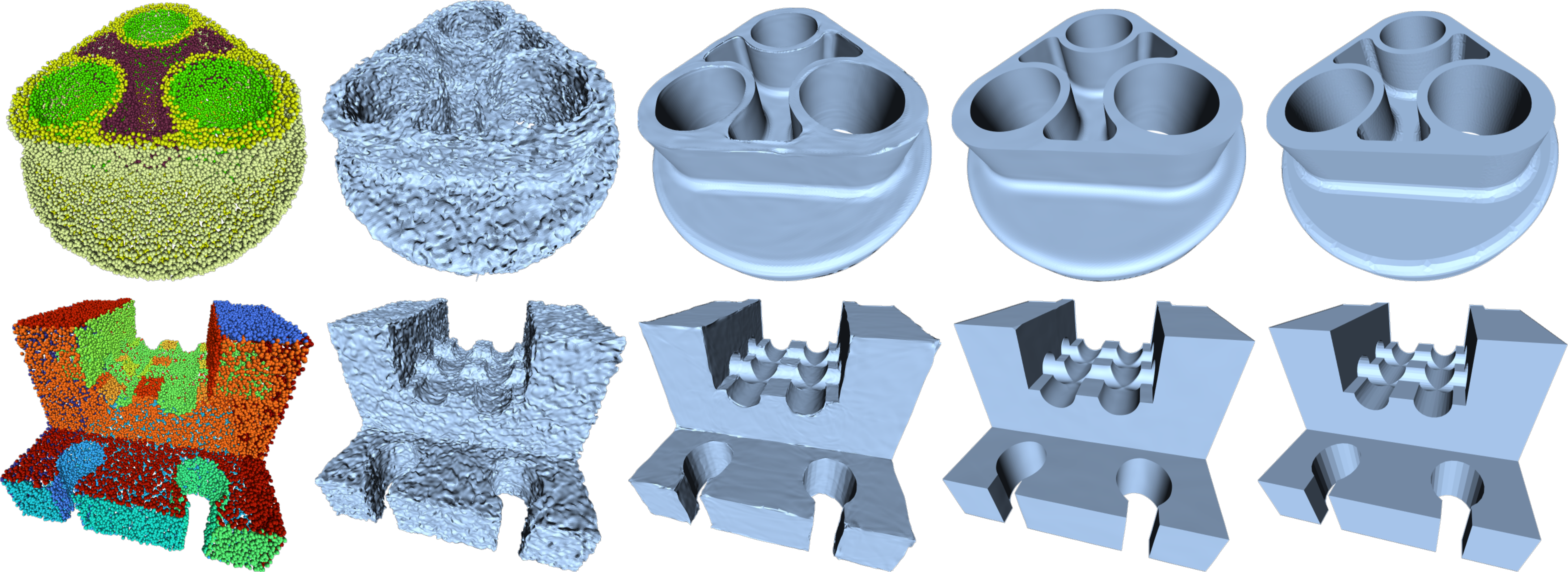}
        \put(3,-3.5){\small Noisy input}
        \put(25,-3.5){\small SIREN}
        \put(47,-3.5){\small IGR}
        \put(67,-3.5){\small Ours}
        \put(82,-3.5){\small Ground truth}
    \end{overpic}\vspace{1pt}
    \caption{NH-Rep conversion from Noisy B-Rep inputs.}
    \label{fig:noise}
\end{figure}

%% file: figures/ourseg.tex
\begin{figure}[t]
    \centering
    \begin{overpic}[width=0.9\columnwidth]{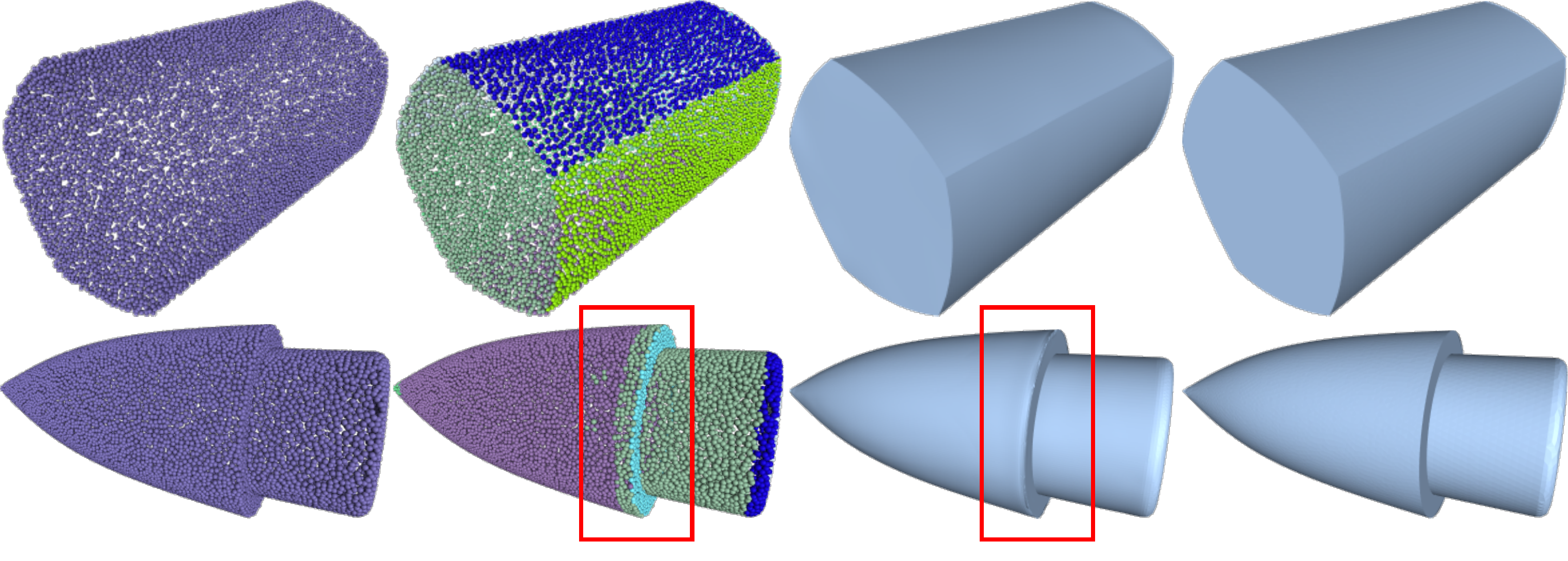}
        \put(9,-1.3){\small Input}
        \put(28,-1.3){\small Segmentation}
        \put(60,-1.3){\small Ours}
        \put(80,-1.3){\small Ground truth}
    \end{overpic}
    \caption{
        Conversion from segmented point clouds to NH-Reps. Some sharp features (bottom) are missed due to incorrect segmentation by \cite{sharma2020parsenet}: the green points occupy two different surface regions.  Each model has 10000 sample points.}
    \label{fig:ourseg} \vspace{-2mm}
\end{figure}

%% file: figures/patchdecom.tex
\begin{figure}[t]
    \centering
    \begin{overpic}[width=1.0\columnwidth]{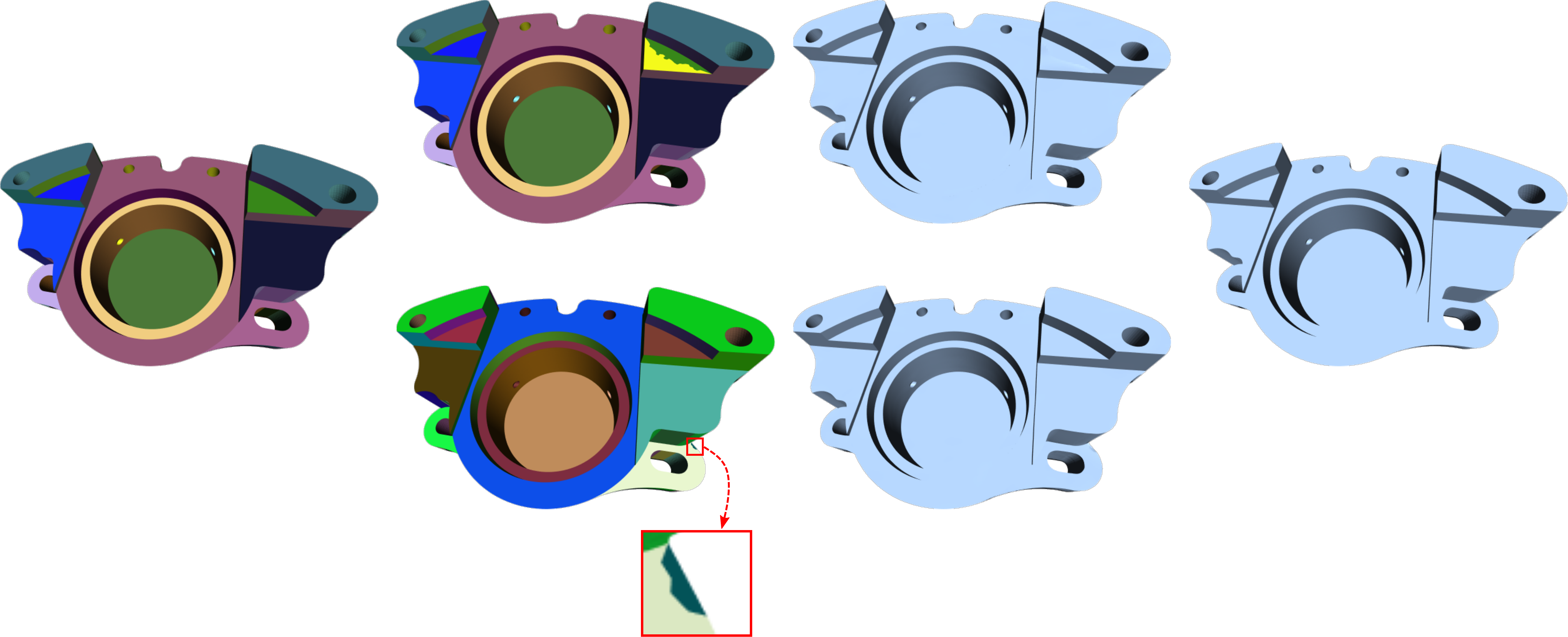}
        \put(9,-1){\small \textbf{(a)}}
        \put(34,-1){\small \textbf{(b)}}
        \put(60,-1){\small \textbf{(c)}}
        \put(86,-1){\small \textbf{(d)}}
    \end{overpic}
    \caption{NH-Rep generated from different patch decompositions. \textbf{(a)}: Input patches of the input model. \textbf{(b)}: Two different decomposed patches due to random initialization. \textbf{(c)}:
        Zero surfaces of the generated NH-Reps; \textbf{(d)}: Ground truth. Patches are rendered in different colors.}
    \label{fig:repair3d} 
\end{figure}

%% file: tables/decomtab.tex
\begin{table}[t]
    \centering \caption{Qualitative evaluation of implicit conversion of the model in \cref{fig:repair3d}.} \label{tab:repair3d}
    \scalebox{0.9}{
        \begin{tabular}{crrrrrrr}
            \toprule
            \thead{Model} & \thead{CD  $\downarrow$} & \thead{HD  $\downarrow$} & \thead{NAE $\downarrow$} & \thead{FCD  $\downarrow$} & \thead{FAE $\downarrow$} & \thead{DE $\downarrow$} & \thead{IoU $\uparrow$} \\
            \midrule
            Top           & 5.87                     & 22.7                     & \ang{5.20}               & 3.11                      & \ang{11.7}               & 0.0531                  & 0.991                  \\
            Bottom        & 5.85                     & 24.9                     & \ang{5.19}               & 2.88                      & \ang{11.4}               & 0.0619                  & 0.993                  \\
            \bottomrule
        \end{tabular}
    }
\end{table}

%% file: figures/scalability.tex
\begin{figure}[t]
    \centering
    \vspace{3pt}
    \begin{overpic}[width=0.95\columnwidth]{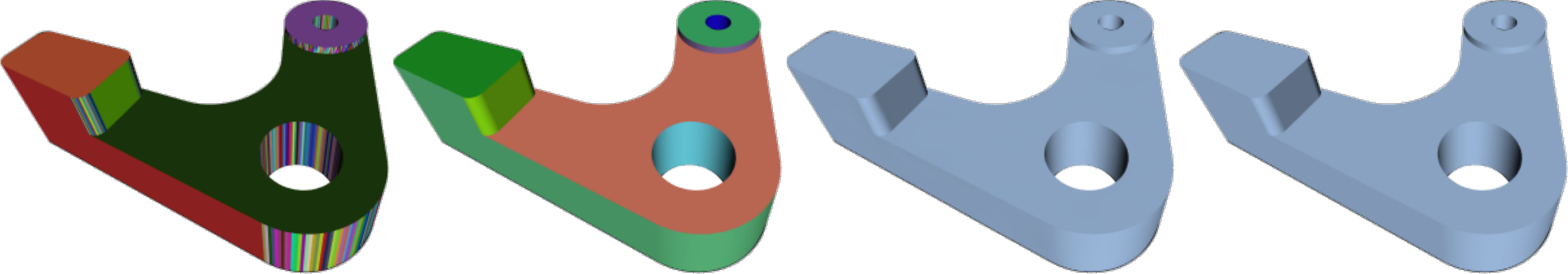}
        \put(2,-3.5){\small Before merging}
        \put(30,-3.5){\small After merging}
        \put(60,-3.5){\small NH-Rep}
        \put(80,-3.5){\small Ground truth}
        \put(0,18){\small {\textbf{\#patches: 868}}}
        \put(28,18){\small {\textbf{\#patches: 11}}}
    \end{overpic}\vspace{1mm}
    \caption{Patch merging for NH-Rep. From left to right: Patches of the input model before and after merging, the zero surface of NH-Rep after patch merging, and the ground truth.}
    \label{fig:scalability} \vspace{-2mm}
\end{figure}

%% file: src/application.tex
\section{Applications} \label{sec:app}
NH-Rep representation can be incorporated with many applications, such as inside / outside query, surface offset, Boolean operations, feature edge blending, and mesh repair.

\paragraph{Fast inside/outside query} Due to the implicit representation of NH-Rep and parallel computation on GPU, it is easy to perform inside/outside queries efficiently. We selected five models from our benchmark dataset, and query the functional values on $2^{17}$ points randomly sampled inside $[-1,1]^3$. The average query time on a single model is \SI{33}{\ms}, which is faster than IGR (\SI{93}{\ms}), and slightly slower than SIREN (\SI{30}{\ms}).

\input{figures/nonzero}
\paragraph{Offset surfaces} The implicit nature of NH-Rep enables us to perform sharp offsets via changing the level sets of $h(\mx)$. \cref{fig:nonzero} illustrates the offset surfaces at iso values: $-0.1, 0, 0.1, 0.2$. NH-Rep maintains sharp features in all offset results. In contrast, IGR cannot model sharp offsets and its offset surfaces are also bumpy with larger iso values; SIREN does not produce plausible offset results.

\input{figures/boolean}
\paragraph{Boolean operations}
NH-Rep supports fast and robust Boolean operations, as no explicit surface intersection is needed. The intersection, union, and complement operated on shapes correspond to $\bm{\max, \min, -}$ operated on implicit functions. \cref{fig:boolean} shows the results of applying Boolean operations on two complicated surfaces represented by NH-Rep.

\input{figures/rfunction}
\paragraph{Feature blending}
NH-Rep can be combined with R-functions \cite{shapiro2007semi} to blend feature edges of solids so that two adjacent patches with sharp features can be joined smoothly. We achieve this goal by replacing the operation $\bm{\max, \min}$ with $\rho$-blending function, which is defined as follows:
\begin{equation}
    B(f, g) = f + g + s\sqrt{f^2 + g^2 + \frac{s_{\rho}(s_{\rho} - |s_{\rho}|)}{8\rho^2} },
\end{equation}
where $f, g$ are the input functions for $\bm{\max}$ or $\bm{\min}$, $\rho$ controls the blend radius and $s_{\rho} = f^2 + g^2 - \rho^2$.  $s=1$ for operation $\bm{\max}$ and $s=-1$ for $\bm{\min}$. Here, since the $\rho$-blending function is bivariable, we rewrite the multivariate Boolean operation of NH-Rep as bivariable operations first, such as $\max\{f_1,f_2,f_3\}=\max\{\max\{f_1,f_2\},f_3\}$, then replace the operator by the $\rho$-blending function.
\cref{fig:rfunction} shows the two examples of feature blending.

\input{figures/meshrepair}
\paragraph{Mesh repairing}
Solid models in B-Rep format are often converted to triangle meshes for different purposes such as file exchange between different 3D software. The conversion may discretize B-Rep patches individually, resulting in mismatching patch edges between adjacent patches, as shown in \cref{fig:meshrepair}-left. Thus, the converted mesh is not watertight and exhibits many visible seams which are not good for downstream tasks such as 3D printing.  These problems can be repaired using NH-Rep. We first segment the mesh based on the existing seams and sharp features and build the adjacent graph on the segmented patches. Here, we regard two disconnected patches with small seams as connected when constructing the Boolean tree. We then sample dense points on the mesh and fit the NH-Rep to them. The formulation of NH-Rep guarantees that a closed mesh with sharp feature preservation can be extracted. \cref{fig:meshrepair} shows the repaired results of two models. Here, we note that our approach is not designed to fix other kinds of mesh imperfections, such as big holes, non-manifold connectivity, and self-intersection.

%% file: figures/nonzero.tex
\begin{figure}[t]
    \centering
    \begin{overpic}[width=0.9\columnwidth]{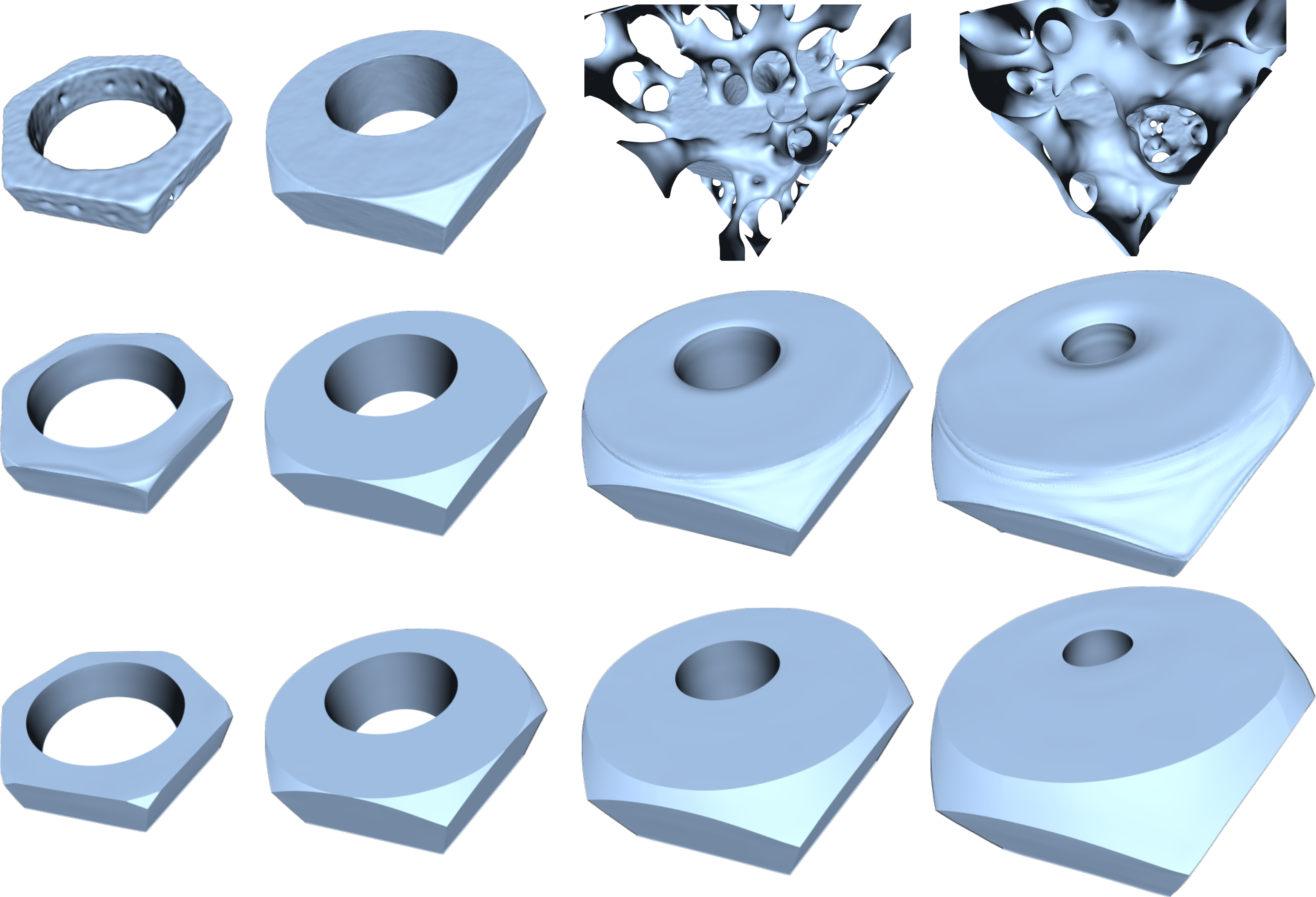}
        \put(0,65.5){\small SIREN}
        \put(0,44){\small IGR}
        \put(0,20){\small Ours}
        \put(3,-3){\small $-0.1$}
        \put(29,-3){\small $0$}
        \put(52,-3){\small $0.1$}
        \put(81,-3){\small $0.2$}
    \end{overpic}
    \caption{Level sets of SIREN (top), IGR (middle), and our method (bottom) with different isovalues: -0.1, 0, 0.1, 0.2, (from left to right).}
    \label{fig:nonzero} \vspace{-2mm}
\end{figure}

%% file: figures/boolean.tex
\begin{figure}[t]
    \centering
    \begin{overpic}[width=0.95\linewidth]{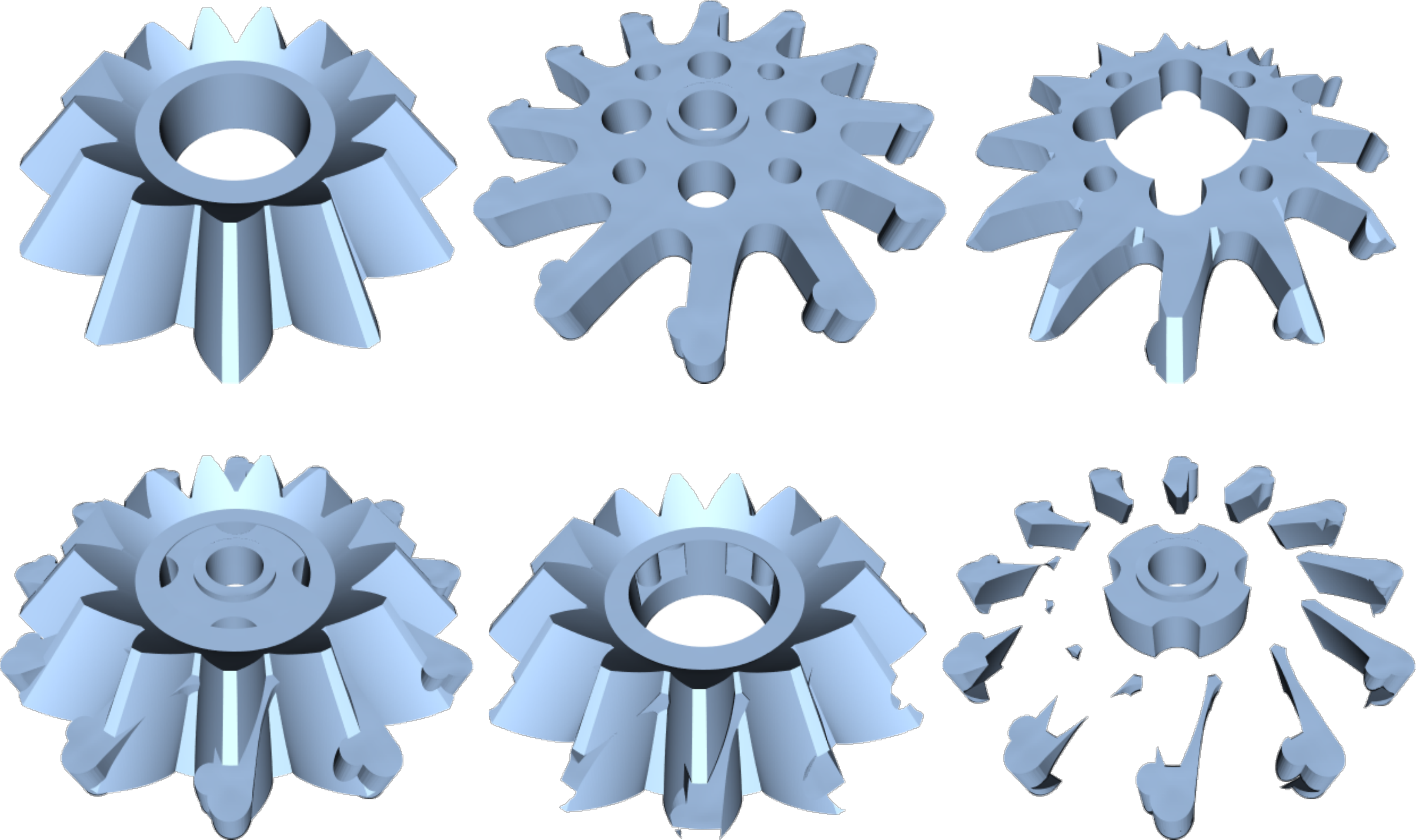}
        \put(15,29){\small $A$}
        \put(49,29){\small $B$}
        \put(79,29){\small $A\cap B$}
        \put(12,-2.5){\small $A\cup B$}
        \put(46,-2.5){\small $A- B$}
        \put(79,-2.5){\small $B- A$}
    \end{overpic} 
    \caption{Zero surfaces of applying boolean operations on two shapes $A$ and $B$ represented by NH-Rep.
    }
    \label{fig:boolean} \vspace{-2mm}
\end{figure}

%% file: figures/rfunction.tex
\begin{figure}[t]
    \centering
    \begin{overpic}[width=0.9\columnwidth]{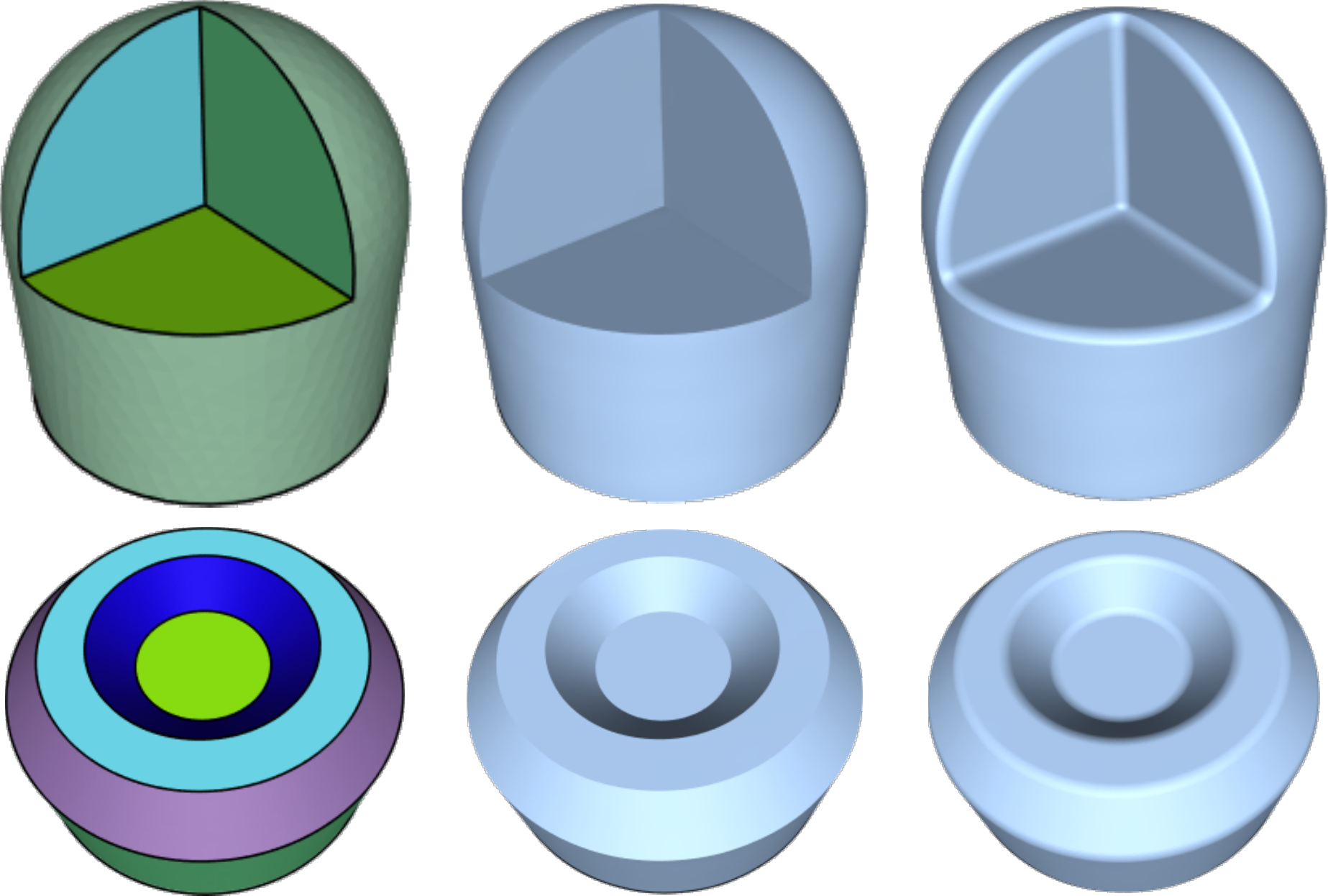}
        \put(7,-2.8){\small Input B-Rep}
        \put(45,-2.8){\small NH-Rep}
        \put(74,-2.8){\small Feature Blending}
    \end{overpic}
    \caption{Feature edge blending of NH-Rep using R-functions.  $\rho$ is set to $0.05$.
    }
    \label{fig:rfunction} \vspace{-2mm}
\end{figure}

%% file: figures/meshrepair.tex
\begin{figure}[t]
    \centering
    \begin{overpic}[width=0.95\columnwidth]{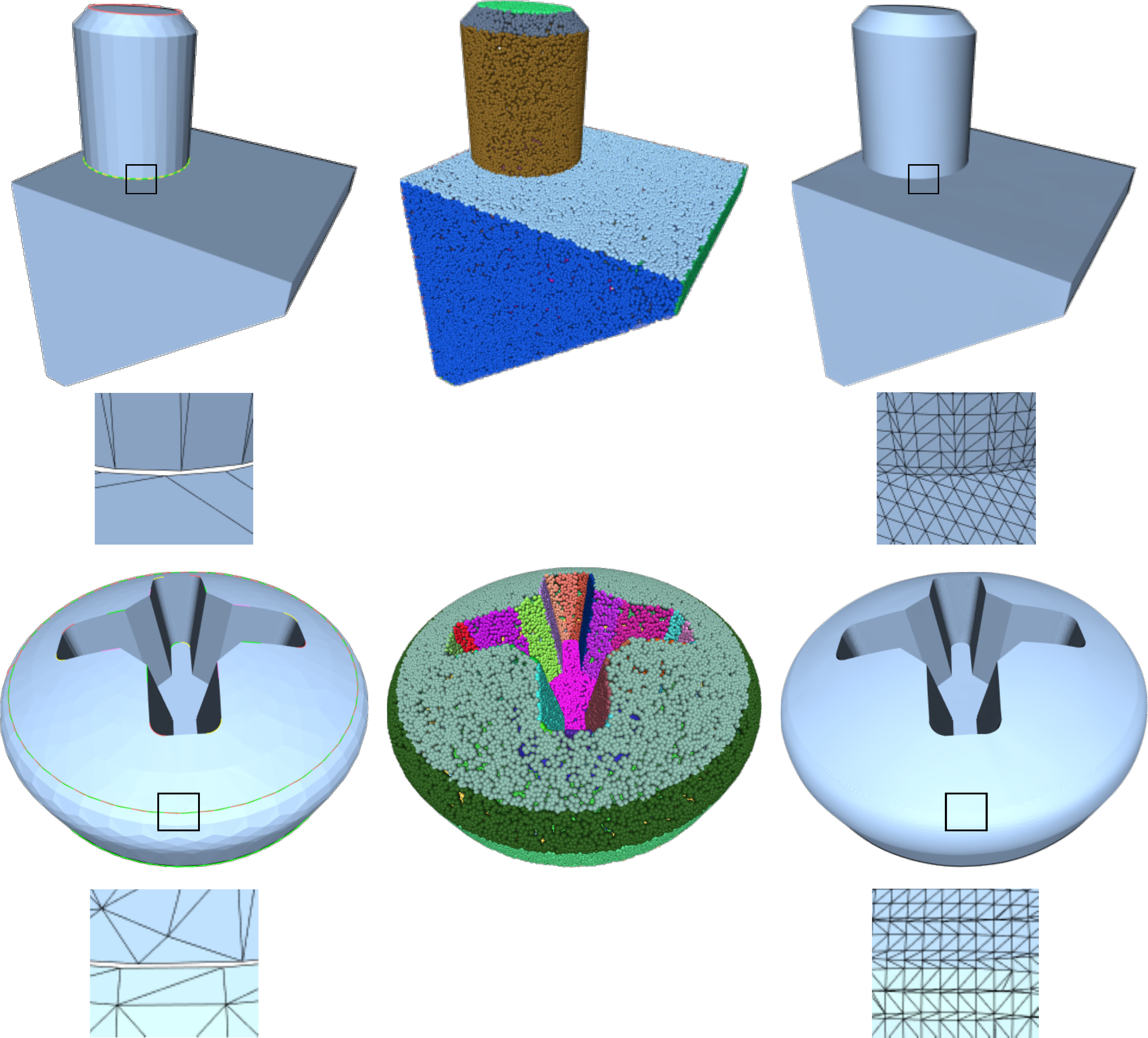}
        \put(6,-3.5){\small Input mesh}
        \put(40,-3.5){\small Sample points}
        \put(74,-3.5){\small Repaired mesh}
    \end{overpic}\vspace{1pt}
    \caption{Mesh repairing. The input meshes are not watertight and contains many small seams (mismatched edges), highlighted with different colors. By sampling dense points on the input mesh (middle) and convert them to NH-Rep by our algorithm, a watertight mesh with feature preserving can be extracted from the zero isosurface of NH-Rep. The zoom-in views highlight the gap before repairing and the seamless results after repairing.}
    \label{fig:meshrepair}
\end{figure}

%% file: src/conclusion.tex
\section{Conclusion} \label{sec:conclusion}
We present neural halfspace representations to convert B-Rep solid models to implicit solids. The efficacy and robustness of our approach are validated through extensive experiments on a large CAD dataset. Compared to other implicit conversion/reconstruction algorithms, our approach offers superior approximation quality and preserves sharp features more faithfully. As demonstrated in \cref{sec:app}, NH-Rep is useful for various applications and it is promising to integrate it with advanced operations applied to function representations~\cite{pasko1995function,shapiro2007semi}, to obtain more shape modeling and optimization capabilities.

Some limitations remain in our work and deserve future development.
First, unlike the traditional CSG representation and the boundary-sampled halfspace representation~\cite{Du2021} that are easy to edit by manipulating their primitives via moving, scaling, rotating, and other simple geometry operations,  it is not intuitive to manipulate the halfspace functions of NH-Rep for shape editing, as its halfspace functions usually do not correspond to simple solids.
Second, the time-consuming learning step hinders interactive shape modeling and implicit conversion. The recent exploration of multilevel and adaptive feature volumes for computing neural implicits~\cite{takikawa2021nglod,mueller2022instant} is promising for accelerating our algorithm. Third, conversion from unsegmented 3D scans to NH-Rep requires reliable segmentation, which is challenging when the input contains large noise, outliers, or insufficient sampling. Lastly, NH-Rep does not support non-manifold B-Rep models. Integrating other representations such as \emph{multiphase implicit functions}~\cite{yuan2012object} into our formulation could be an interesting research direction.

%% file: src/appendix.tex
\appendix

\section{A failure case of boundary-sampled halfspace}\label{appendix:bsh}
\input{figures/bsh}
The limitation of BSH~\cite{Du2021} in handling patches tangentially contacted occurs often in CAD B-Rep models. \cref{fig:bsh-failure} illustrates a simple failure case. The input B-Rep model consists of two cylindrical halfspaces and four planar halfspaces. Each cylindrical patch and its adjacent planar patch are contacted tangentially. The BSH algorithm cannot produce the correct result, while our approach works well.

\input{src/proof}

\section{Evaluation metrics}\label{appendix:metric}
We denote $M_e$ as the extracted zero isosurface from any method, and $M_g$ as the boundary surface of B-Rep. $\mathcal{P}_e$ and $\mathcal{P}_g$ are randomly sampled points on $M_e$ and $M_g$, and $\mathcal{F}_e$ and $\mathcal{F}_g$ are randomly sampled points on the feature edges of $M_e$ and $M_g$, respectively. For a point set $\mathcal{H}$ and a query point $\mx \in \mathbb{R}^3$, we define the operation $Q(\mx, \mathcal{H}) = \argmin_{\my \in \mathcal{H}} \| \mx - \my\|$ that returns the nearest point to $\mx$ in $\mathcal{H}$, the operator $ \mn(\mx)$ that returns the surface normal at $\mx$, and the operator $\theta(\mx)$ that returns the dihedral angle at a feature point $\mx$. The evaluation metrics~\cref{subsec:setup} are listed below.

\begin{align*}
    \texttt{CD}(M_e, M_g)   = & \frac{1}{2|\mathcal{P}_e|} \sum_{\mp \in \mathcal{P}_e} \| \mp - Q(\mp, \mathcal{P}_g) \| + \frac{1}{2|\mathcal{P}_g|} \sum_{\mp \in \mathcal{P}_g} \| \mp - Q(\mp, \mathcal{P}_e) \|; \\
    \texttt{HD}(M_e, M_g)   = & \max\{ \max_{\mp \in \mathcal{P}_e} \| \mp - Q(\mp, \mathcal{P}_g) \|,  \max_{\mp \in \mathcal{P}_g} \| \mp - Q(\mp, \mathcal{P}_e) \| \};                                             \\
    \texttt{NAE}(M_e, M_g)  = & \frac{\ang{180}}{2|\mathcal{P}_e|\pi} \sum_{\mp \in \mathcal{P}_e} \arccos \bigl(\mn(\mp) \cdot \mn(Q(\mp, \mathcal{P}_g))\bigr)                                                       \\
                              & + \frac{\ang{180}}{2|\mathcal{P}_g|\pi} \sum_{\mp \in \mathcal{P}_g} \arccos \bigl(\mn(\mp) \cdot \mn(Q(\mp, \mathcal{P}_e))\bigr);                                                    \\
    \texttt{FCD}(M_e, M_g)  = & \frac{1}{2|\mathcal{F}_e|} \sum_{\mp \in \mathcal{F}_e} \| \mp - Q(\mp, \mathcal{F}_g) \| + \frac{1}{2|\mathcal{F}_g|} \sum_{\mp \in \mathcal{F}_g} \| \mp - Q(\mp, \mathcal{F}_e) \|; \\
    \texttt{FAE}(M_e, M_g)  = & \frac{1}{2|\mathcal{F}_e|} \sum_{\mp \in \mathcal{F}_e} \| \theta(\mp) - \theta(Q(\mp, \mathcal{F}_g)) \|                                                                              \\
                              & + \frac{1}{2|\mathcal{F}_g|} \sum_{\mp \in \mathcal{F}_g} \| \theta(\mp) - \theta(Q(\mp, \mathcal{F}_e)) \|.
\end{align*}

For a B-Rep solid, we compute its ``ground-truth'' signed distance field $\mathcal{F}_g$ based on the accompanied triangle mesh in the ABC dataset, we measure the approximation error between the learned function $\mathcal{F}_e$ to $\mathcal{F}_g$ as follows:
\begin{equation*}
    \texttt{DE}(\mathcal{F}_e, \mathcal{F}_g)  =  \frac{1}{|G|}\sum_{\mp \in G} \frac{|\mathcal{F}_g(\mp) - f(\mp)|}{|\mathcal{F}_g(\mp)| + \delta}.
\end{equation*}
Here $G$ is a set of points randomly sampled in $[-1,1]^3$, $|G|=2^{17}$ and $\delta > 0$ is a tiny value to avoid zero division and is set to $10^{-9}$.

\textbf{IoU} is the volumetric intersection of $M_g$ and $M_e$ divided by their volume union. $2^{17}$ points are randomly sampled in $[-1,1]^3$ and their occupancy values are evaluated for computing \textbf{IoU}.

%% file: figures/bsh.tex
\begin{figure}[t]
    \centering
    \includegraphics[width=1\columnwidth]{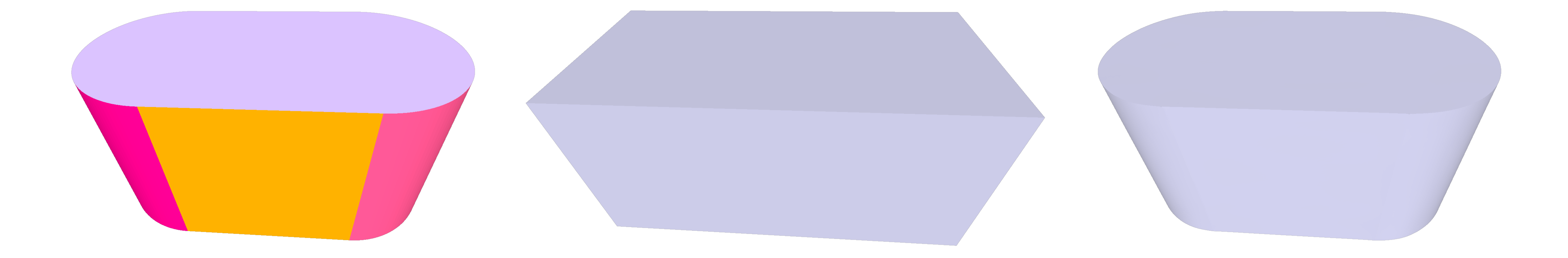}
    \caption{Left: The input B-Rep.  Middle: The reconstruction result by BSH~\cite{Du2021}. Right: The zero isosurface of our NH-Rep.} \label{fig:bsh-failure} \vspace{-2mm}
\end{figure}

%% file: src/proof.tex
\section{Existence of Implicit Functions} \label{appendix:proof}
In this section, we prove that for a decomposed patch set $\{\ms_1, \ldots, \ms_n\}$ of a B-Rep solid $\mS$, there exist a series of implicit functions $\{f_1,  \ldots, f_n\}$ such that their composite function $h$ satisfies \cref{eq:classification}.

It is known that an implicit function $F$ can be induced from an oriented manifold surface $\mZ$ (open or closed without self-intersection), with the condition that $\mZ$ belongs to the zero isosurface of $F$. To prove the existence of $\{f_i\}_{i=1}^n$,  it is equivalent to proving that there exists a set of orientable manifold surfaces: $\{\mZ_i\}_{i=1}^n$ which can induce $\{f_i\}_{i=1}^n$ to satisfy \cref{eq:classification}.

We prove the existence of $\{\mZ_i\}_{i=1}^n$ in a constructive way by creating $\mZ_i$ from top to bottom along the Boolean tree, as follows.
Without loss of generality, we show how to create relevant $\mZ_i$s on an arbitrary Boolean tree node. We assume that a tree node $r$ has child patches $\mQ_1,\ldots,\mQ_m$, child nodes $r_1, \ldots,r_d$. Here $\mQ_i, i=1,\ldots,m$ are a subset of $\{\ms_k\}_{k=1}^n$. We denote $\mQ_{m+i}$ as the union of patches contained in $r_i$ and its descendants. 
In the following construction process, we assume $\bm\op(r) = \bm\max$. When $\bm\op(r) = \bm\min$, the process is symmetric.

\begin{figure}[t]
    \centering
    \begin{overpic}[width=0.8\columnwidth]{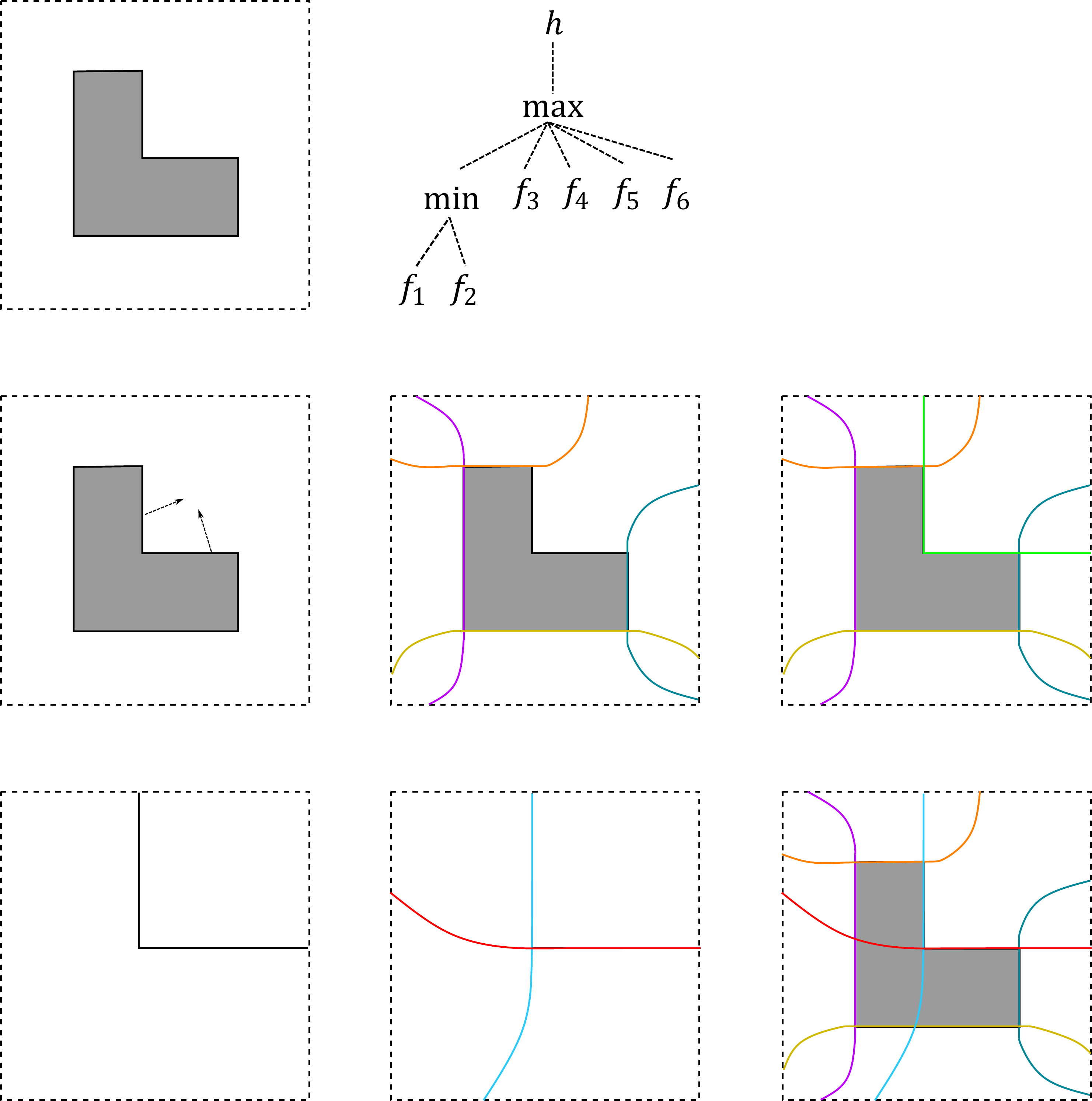}
        \put(12,69){\small \textbf{(a)}}
        \put(48,69){\small \textbf{(b)}}
        \put(12,33){\small \textbf{(c)}}
        \put(48,33){\small \textbf{(d)}}
        \put(83,33){\small \textbf{(e)}}
        \put(12,-3){\small \textbf{(f)}}
        \put(48,-3){\small \textbf{(g)}}
        \put(83,-3){\small \textbf{(h)}}

        \put(12,76){\scriptsize \textbf{$\ms_4$}}
        \put(13,91){\scriptsize \textbf{$\ms_1$}}
        \put(17,87){\scriptsize \textbf{$\ms_2$}}
        \put(8,95){\scriptsize \textbf{$\ms_6$}}
        \put(22,82){\scriptsize \textbf{$\ms_3$}}
        \put(3,86){\scriptsize \textbf{$\ms_5$}}

        \put(12,40){\scriptsize \textbf{$\mQ_3$}}
        \put(17,55){\scriptsize \textbf{$\mQ_5$}}
        \put(8,59){\scriptsize \textbf{$\mQ_1$}}
        \put(22,46){\scriptsize \textbf{$\mQ_2$}}
        \put(2,50){\scriptsize \textbf{$\mQ_4$}}

        \put(48,40){\scriptsize \textcolor[RGB]{209,185,0}{\textbf{$z_3$}}}
        \put(44,59){\scriptsize \textcolor[RGB]{255,128,0}{\textbf{$z_1$}}}
        \put(57,46){\scriptsize \textcolor[RGB]{0,135,152}{\textbf{$z_2$}}}
        \put(38,50){\scriptsize \textcolor[RGB]{191,0,255}{\textbf{$z_4$}}}

        \put(83,40){\scriptsize \textcolor[RGB]{209,185,0}{\textbf{$z_3$}}}
        \put(79,59){\scriptsize \textcolor[RGB]{255,128,0}{\textbf{$z_1$}}}
        \put(93,46){\scriptsize \textcolor[RGB]{0,135,152}{\textbf{$z_2$}}}
        \put(73,50){\scriptsize \textcolor[RGB]{191,0,255}{\textbf{$z_4$}}}
        \put(88,55){\scriptsize \textcolor[RGB]{0,255,0}{\textbf{$z_5$}}}

        \put(18,15.5){\scriptsize \textbf{$\mQ'_2$}}
        \put(13,22){\scriptsize   \textbf{$\mQ'_1$}}

        \put(54,15.5){\scriptsize \textcolor[RGB]{255,0,0}{\textbf{$z'_2$}}}
        \put(49,22){\scriptsize   \textcolor[RGB]{39,204,255}{\textbf{$z'_1$}}}

    \end{overpic} 
    \caption{ Construction of extended curves on a 2D model. The extended curves are trimmed by the bounding box for better illustration.
    }
    \label{fig:genus0} \vspace{-6mm}
\end{figure}

\begin{figure}[t]
    \centering
    \begin{overpic}[width=0.6\columnwidth]{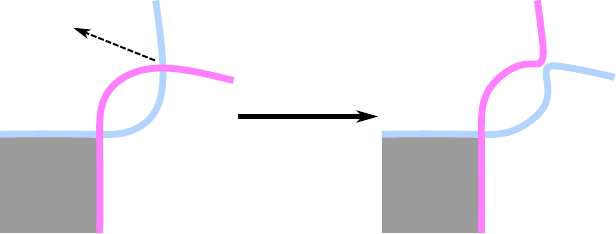}
        \put(-7,34){\scriptsize unwanted}
        \put(-8,30){\scriptsize intersection}
    \end{overpic}
    \caption{Removal of unwanted intersection by exchanging and smoothing when constructing 2D zero curves.}
    \label{fig:intersection}
\end{figure}

We process $\mQ_{s}, s=1,\ldots,m+d$ in sequence as follows. If $\mQ_s$ lies on the outer boundary of the shape (\eg $\mathcal{Q}_1$ in \cref{fig:genus0}),  we can extend $\mQ_s$ from the boundary of $\mQ_s$ to form an open orientable and non-self-intersecting surface $\mZ_{\mQ_s}$. The extension of $\mZ_s$ should also avoid intersecting with other $\mQ_i$'s except at the boundary of $\mQ_s$. This can be achieved because $\mQ_s$ lies on the boundary of a connected region and the boundary feature curves of $\mQ_s$ are all convex. All constructed zero surfaces have no intersection with the boundary surface of $\mS$ except on feature curves (or corners in 2D). Next, we need to remove all the other intersections of $\mZ_i$'s outside $\mS$, otherwise, the composited halfspace will have outlier regions. Unwanted intersections can be removed by exchanging and smoothing the intersected parts, as shown in \cref{fig:intersection}.

\begin{figure}[t]
    \centering
    \begin{overpic}[width=0.8\columnwidth]{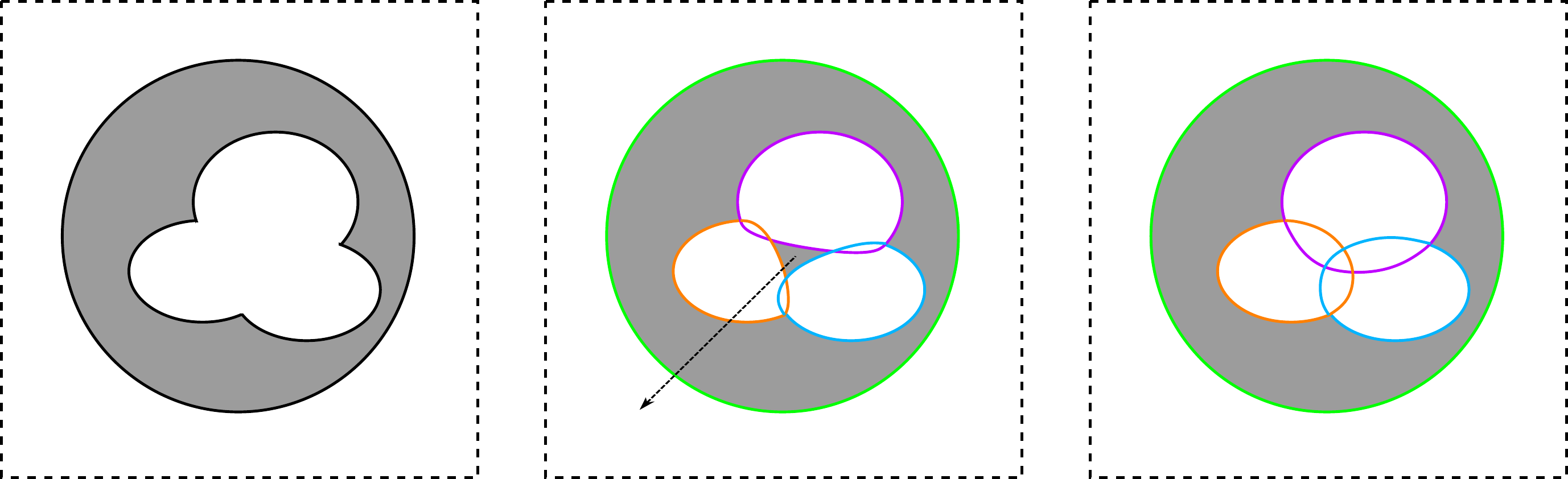}
        \put(13,-3){\scriptsize \textbf{(a)}}
        \put(49,-3){\scriptsize \textbf{(b)}}
        \put(83,-3){\scriptsize \textbf{(c)}}

        \put(4.5,24){\scriptsize \textbf{$\mQ_1$}}
        \put(10,11){\scriptsize \textbf{$\mQ_2$}}
        \put(16,18.5){\scriptsize \textbf{$\mQ_3$}}
        \put(19.8,10.5){\scriptsize \textbf{$\mQ_4$}}

        \put(39.5,24){\scriptsize \textcolor[RGB]{0,255,0}{\textbf{$z_1$}}}
        \put(44.5,12){\scriptsize \textcolor[RGB]{255,128,0}{\textbf{$z_2$}}}
        \put(51,18.5){\scriptsize \textcolor[RGB]{191,0,255}{\textbf{$z_3$}}}
        \put(54.8,10.5){\scriptsize \textcolor[RGB]{0,180,254}{\textbf{$z_4$}}}

        \put(35.5,2){\scriptsize Interior outlier}

        \put(74.5,24){\scriptsize \textcolor[RGB]{0,255,0}{\textbf{$z_1$}}}
        \put(79.5,12){\scriptsize \textcolor[RGB]{255,128,0}{\textbf{$z_2$}}}
        \put(86,18.5){\scriptsize \textcolor[RGB]{191,0,255}{\textbf{$z_3$}}}
        \put(89.8,10.5){\scriptsize \textcolor[RGB]{0,180,254}{\textbf{$z_4$}}}

    \end{overpic}
    \caption{Construction of extended curves on a 2D model with genus 1. \textbf{(a)}: Input model with 4 patches. \textbf{(b)}: Improper zero surfaces lead to interior outliers. \textbf{(c)}: After expanding the zero surfaces, interior outliers can be removed.
    }
    \label{fig:genus1} \vspace{-2mm}
\end{figure}

If $\mQ_s$ lies on the boundary of a cavity region of the solid (\eg $\mathcal{Q}_2$ in \cref{fig:genus1}), the zero surface can be constructed in a similar way, but the resulting surface would be a closed one.  If the composited halfspace has an interior outlier region within the cavity, the zero surfaces should be expanded to cover the outlier region, as shown in \cref{fig:genus1}(c).

After constructing zero surfaces for all $\mQ_i$'s, the subpatches of $\mQ_{m + i}, i=1,\ldots,d$ should also be replaced with their extended versions. Suppose that $\mR_j$'s are the subpatches of $\mQ_{m + i}$, then $\mR_j$ should be extended along $\partial\mR_j \cap \partial \mQ_{m + i}$, so that the union of all extended $\mR_j$'s is $\mZ_{\mQ_{m + i}}$.
In this way, the generated zero surfaces in the child nodes of $r$ follow that of $r$.

The above procedure is executed from top to bottom; we can obtain a series of $\{\mZ_i\}_{i=1}^n$ at each leaf node. From $\{\mZ_i\}_{i=1}^n$, we can induce a set of implicit functions: $\{f_i\}_{i=1}^n$. Due to the construction process and the above properties, it is easy to verify that \cref{eq:classification} can be fulfilled.

In \cref{fig:genus0} and \cref{fig:genus1}, we illustrate the concept of surface extension in two 2D examples.
\cref{fig:genus0}-(a) presents an input shape with six decomposed patches $\ms_1, \ldots, \ms_6$. Its Boolean tree is illustrated in \cref{fig:genus0}-(b), $f_i$ corresponds to $\ms_i$. \cref{fig:genus0}-(c) is the grouped patches at the second layer of the Boolean tree, where $\mQ_5$ is the union of $\ms_1$ and $\ms_2$. The extended curves $\mZ_{\mQ_1}, \ldots, \mZ_{\mQ_4}$ are illustrated in \cref{fig:genus0}-(c), denoted by $z_1, \ldots, z_4$.  $\mQ_5$ is processed at the child node (\cref{fig:genus0}-(f)) and the corresponding extended curves are $z'_1$ and $z'_2$ shown in \cref{fig:genus0}-(g). By applying the union operation (corresponding to $\min$ operation), $z_5$ is constructed as shown in \cref{fig:genus0}-(e). In \cref{fig:genus0}-(h), all extended curves are plotted.  \cref{fig:genus1}(a) shows another example whose input is with genus 1. The extended curves of $\mQ_1,\mQ_2$, $\mQ_3$ and $\mQ_4$ are closed curves as depicted in \cref{fig:genus1}-(c).

Here, note that a dedicated and complicated algorithm is needed to implement the above surface extension procedure, and we turn to train a neural network to compute the appropriate implicit functions directly.

